\documentclass[longauth]{aa}  
\usepackage{graphicx}
\usepackage{hyperref}
\usepackage{euclid}
\usepackage{bm}
\usepackage{comment}
\usepackage[normalem]{ulem}

\usepackage[dvipsnames]{xcolor}
\usepackage[T1]{fontenc} 
\usepackage{braket}

\usepackage[nameinlink,noabbrev]{cleveref}
\Crefname{equation}{Eq.}{Eqs.}
\Crefname{section}{Sect.}{Sects.}
\Crefname{figure}{Fig.}{Figs.}
\crefname{equation}{Equation}{Equations}
\crefname{section}{Section}{Sections}
\crefname{figure}{Figure}{Figures}

\defcitealias{ISTFpaper}{EC20}
\begin{document} 

\title{\Euclid: constraining dark energy coupled to electromagnetism using astrophysical and laboratory data\thanks{This paper is published on behalf of the Euclid Consortium.}}

\author{M.~Martinelli$^{1}$\thanks{\email{matteo.martinelli@uam.es}}, C.J.A.P.~Martins$^{2,3}$, S.~Nesseris$^{1}$, I.~Tutusaus$^{4,5,6}$, A.~Blanchard$^{5}$, S.~Camera$^{7,8,9}$, C.~Carbone$^{10}$, S.~Casas$^{11}$, V.~Pettorino$^{11}$, Z.~Sakr$^{5,12}$, V.~Yankelevich$^{13}$, D.~Sapone$^{14}$, A.~Amara$^{15}$, N.~Auricchio$^{16}$, C.~Bodendorf$^{17}$, D.~Bonino$^{9}$, E.~Branchini$^{18,19}$, V.~Capobianco$^{9}$, J.~Carretero$^{20}$, M.~Castellano$^{21}$, S.~Cavuoti$^{22,23,24}$, A.~Cimatti$^{25,26}$, R.~Cledassou$^{27,28}$, L.~Corcione$^{9}$, A.~Costille$^{29}$, H.~Degaudenzi$^{30}$, M.~Douspis$^{31}$, F.~Dubath$^{30}$, S.~Dusini$^{32}$, A.~Ealet$^{33}$, S.~Ferriol$^{33}$, M.~Frailis$^{34}$, E.~Franceschi$^{16}$, B.~Garilli$^{10}$, C.~Giocoli$^{35,36}$, A.~Grazian$^{37}$, F.~Grupp$^{17,38}$, S.V.H.~Haugan$^{39}$, W.~Holmes$^{40}$, F.~Hormuth$^{41,42}$, K.~Jahnke$^{42}$, A.~Kiessling$^{40}$, M.~K\"ummel$^{38}$, M.~Kunz$^{43}$, H.~Kurki-Suonio$^{44}$, S.~Ligori$^{9}$, P.B.~Lilje$^{39}$, I.~Lloro$^{45}$, O.~Mansutti$^{34}$, O.~Marggraf$^{46}$, K.~Markovic$^{40}$, R.~Massey$^{47}$, M.~Meneghetti$^{16,48}$, G.~Meylan$^{49}$, L.~Moscardini$^{16,25,50}$, S.M.~Niemi$^{51}$, C.~Padilla$^{20}$, S.~Paltani$^{30}$, F.~Pasian$^{34}$, K.~Pedersen$^{52}$, S.~Pires$^{11}$, M.~Poncet$^{28}$, L.~Popa$^{53}$, F.~Raison$^{17}$, R.~Rebolo$^{54,55}$, J.~Rhodes$^{40}$, M.~Roncarelli$^{16,25}$, E.~Rossetti$^{25}$, R.~Saglia$^{17,38}$, A.~Secroun$^{56}$, G.~Seidel$^{42}$, S.~Serrano$^{4,6}$, C.~Sirignano$^{32,57}$, G.~Sirri$^{50}$, J.-L.~Starck$^{11}$, D.~Tavagnacco$^{34}$, A.N.~Taylor$^{58}$, I.~Tereno$^{59,60}$, R.~Toledo-Moreo$^{61}$, L.~Valenziano$^{16,50}$, Y.~Wang$^{62}$, G.~Zamorani$^{16}$, J.~Zoubian$^{56}$, M.~Baldi$^{16,50,63}$, M.~Brescia$^{24}$, G.~Congedo$^{58}$, L.~Conversi$^{64,65}$, Y.~Copin$^{33}$, G.~Fabbian$^{66}$, R.~Farinelli$^{67}$, E.~Medinaceli$^{16}$, S.~Mei$^{68}$, G.~Polenta$^{69}$, E.~Romelli$^{34}$, T.~Vassallo$^{38}$}

\institute{$^{1}$ Instituto de F\'isica Te\'orica UAM-CSIC, Campus de Cantoblanco, E-28049 Madrid, Spain\\
$^{2}$ Centro de Astrof\'{\i}sica da Universidade do Porto, Rua das Estrelas, 4150-762 Porto, Portugal\\
$^{3}$ Instituto de Astrof\'isica e Ci\^encias do Espa\c{c}o, Universidade do Porto, CAUP, Rua das Estrelas, PT4150-762 Porto, Portugal\\
$^{4}$ Institute of Space Sciences (ICE, CSIC), Campus UAB, Carrer de Can Magrans, s/n, 08193 Barcelona, Spain\\
$^{5}$ Institut de Recherche en Astrophysique et Plan\'etologie (IRAP), Universit\'e de Toulouse, CNRS, UPS, CNES, 14 Av. Edouard Belin, F-31400 Toulouse, France\\
$^{6}$ Institut d’Estudis Espacials de Catalunya (IEEC), Carrer Gran Capit\'a 2-4, 08034 Barcelona, Spain\\
$^{7}$ INFN-Sezione di Torino, Via P. Giuria 1, I-10125 Torino, Italy\\
$^{8}$ Dipartimento di Fisica, Universit\'a degli Studi di Torino, Via P. Giuria 1, I-10125 Torino, Italy\\
$^{9}$ INAF-Osservatorio Astrofisico di Torino, Via Osservatorio 20, I-10025 Pino Torinese (TO), Italy\\
$^{10}$ INAF-IASF Milano, Via Alfonso Corti 12, I-20133 Milano, Italy\\
$^{11}$ AIM, CEA, CNRS, Universit\'{e} Paris-Saclay, Universit\'{e} de Paris, F-91191 Gif-sur-Yvette, France\\
$^{12}$ Universit\'e St Joseph; UR EGFEM, Faculty of Sciences, Beirut, Lebanon\\
$^{13}$ Astrophysics Research Institute, Liverpool John Moores University, 146 Brownlow Hill, Liverpool L3 5RF, UK\\
$^{14}$ Departamento de F\'isica, FCFM, Universidad de Chile, Blanco Encalada 2008, Santiago, Chile\\
$^{15}$ Institute of Cosmology and Gravitation, University of Portsmouth, Portsmouth PO1 3FX, UK\\
$^{16}$ INAF-Osservatorio di Astrofisica e Scienza dello Spazio di Bologna, Via Piero Gobetti 93/3, I-40129 Bologna, Italy\\
$^{17}$ Max Planck Institute for Extraterrestrial Physics, Giessenbachstr. 1, D-85748 Garching, Germany\\
$^{18}$ INFN-Sezione di Roma Tre, Via della Vasca Navale 84, I-00146, Roma, Italy\\
$^{19}$ Department of Mathematics and Physics, Roma Tre University, Via della Vasca Navale 84, I-00146 Rome, Italy\\
$^{20}$ Institut de F\'{i}sica d’Altes Energies (IFAE), The Barcelona Institute of Science and Technology, Campus UAB, 08193 Bellaterra (Barcelona), Spain\\
$^{21}$ INAF-Osservatorio Astronomico di Roma, Via Frascati 33, I-00078 Monteporzio Catone, Italy\\
$^{22}$ Department of Physics "E. Pancini", University Federico II, Via Cinthia 6, I-80126, Napoli, Italy\\
$^{23}$ INFN section of Naples, Via Cinthia 6, I-80126, Napoli, Italy\\
$^{24}$ INAF-Osservatorio Astronomico di Capodimonte, Via Moiariello 16, I-80131 Napoli, Italy\\
$^{25}$ Dipartimento di Fisica e Astronomia “Augusto Righi” - Alma Mater Studiorum Università di Bologna, via Piero Gobetti 93/2, I-40129 Bologna, Italy\\
$^{26}$ INAF-Osservatorio Astrofisico di Arcetri, Largo E. Fermi 5, I-50125, Firenze, Italy\\
$^{27}$ Institut national de physique nucl\'eaire et de physique des particules, 3 rue Michel-Ange, 75794 Paris C\'edex 16, France\\
$^{28}$ Centre National d'Etudes Spatiales, Toulouse, France\\
$^{29}$ Aix-Marseille Univ, CNRS, CNES, LAM, Marseille, France\\
$^{30}$ Department of Astronomy, University of Geneva, ch. d\'Ecogia 16, CH-1290 Versoix, Switzerland\\
$^{31}$ Universit\'e Paris-Saclay, CNRS, Institut d'astrophysique spatiale, 91405, Orsay, France\\
$^{32}$ INFN-Padova, Via Marzolo 8, I-35131 Padova, Italy\\
$^{33}$ Univ Lyon, Univ Claude Bernard Lyon 1, CNRS/IN2P3, IP2I Lyon, UMR 5822, F-69622, Villeurbanne, France\\
$^{34}$ INAF-Osservatorio Astronomico di Trieste, Via G. B. Tiepolo 11, I-34131 Trieste, Italy\\
$^{35}$ Istituto Nazionale di Astrofisica (INAF) - Osservatorio di Astrofisica e Scienza dello Spazio (OAS), Via Gobetti 93/3, I-40127 Bologna, Italy\\
$^{36}$ Istituto Nazionale di Fisica Nucleare, Sezione di Bologna, Via Irnerio 46, I-40126 Bologna, Italy\\
$^{37}$ INAF-Osservatorio Astronomico di Padova, Via dell'Osservatorio 5, I-35122 Padova, Italy\\
$^{38}$ Universit\"ats-Sternwarte M\"unchen, Fakult\"at f\"ur Physik, Ludwig-Maximilians-Universit\"at M\"unchen, Scheinerstrasse 1, 81679 M\"unchen, Germany\\
$^{39}$ Institute of Theoretical Astrophysics, University of Oslo, P.O. Box 1029 Blindern, N-0315 Oslo, Norway\\
$^{40}$ Jet Propulsion Laboratory, California Institute of Technology, 4800 Oak Grove Drive, Pasadena, CA, 91109, USA\\
$^{41}$ von Hoerner \& Sulger GmbH, Schlo{\ss}Platz 8, D-68723 Schwetzingen, Germany\\
$^{42}$ Max-Planck-Institut f\"ur Astronomie, K\"onigstuhl 17, D-69117 Heidelberg, Germany\\
$^{43}$ Universit\'e de Gen\`eve, D\'epartement de Physique Th\'eorique and Centre for Astroparticle Physics, 24 quai Ernest-Ansermet, CH-1211 Gen\`eve 4, Switzerland\\
$^{44}$ Department of Physics and Helsinki Institute of Physics, Gustaf H\"allstr\"omin katu 2, 00014 University of Helsinki, Finland\\
$^{45}$ NOVA optical infrared instrumentation group at ASTRON, Oude Hoogeveensedijk 4, 7991PD, Dwingeloo, The Netherlands\\
$^{46}$ Argelander-Institut f\"ur Astronomie, Universit\"at Bonn, Auf dem H\"ugel 71, 53121 Bonn, Germany\\
$^{47}$ Institute for Computational Cosmology, Department of Physics, Durham University, South Road, Durham, DH1 3LE, UK\\
$^{48}$ INFN-Bologna, Via Irnerio 46, I-40126 Bologna, Italy\\
$^{49}$ Observatoire de Sauverny, Ecole Polytechnique F\'ed\'erale de Lau- sanne, CH-1290 Versoix, Switzerland\\
$^{50}$ INFN-Sezione di Bologna, Viale Berti Pichat 6/2, I-40127 Bologna, Italy\\
$^{51}$ European Space Agency/ESTEC, Keplerlaan 1, 2201 AZ Noordwijk, The Netherlands\\
$^{52}$ Department of Physics and Astronomy, University of Aarhus, Ny Munkegade 120, DK–8000 Aarhus C, Denmark\\
$^{53}$ Institute of Space Science, Bucharest, Ro-077125, Romania\\
$^{54}$ Departamento de Astrof\'{i}sica, Universidad de La Laguna, E-38206, La Laguna, Tenerife, Spain\\
$^{55}$ Instituto de Astrof\'{i}sica de Canarias. Calle V\'{i}a L\`{a}ctea s/n, 38204, San Crist\'{o}bal de la Laguna, Tenerife, Spain\\
$^{56}$ Aix-Marseille Univ, CNRS/IN2P3, CPPM, Marseille, France\\
$^{57}$ Dipartimento di Fisica e Astronomia “G.Galilei", Universit\'a di Padova, Via Marzolo 8, I-35131 Padova, Italy\\
$^{58}$ Institute for Astronomy, University of Edinburgh, Royal Observatory, Blackford Hill, Edinburgh EH9 3HJ, UK\\
$^{59}$ Instituto de Astrof\'isica e Ci\^encias do Espa\c{c}o, Faculdade de Ci\^encias, Universidade de Lisboa, Tapada da Ajuda, PT-1349-018 Lisboa, Portugal\\
$^{60}$ Departamento de F\'isica, Faculdade de Ci\^encias, Universidade de Lisboa, Edif\'icio C8, Campo Grande, PT1749-016 Lisboa, Portugal\\
$^{61}$ Universidad Polit\'ecnica de Cartagena, Departamento de Electr\'onica y Tecnolog\'ia de Computadoras, 30202 Cartagena, Spain\\
$^{62}$ Infrared Processing and Analysis Center, California Institute of Technology, Pasadena, CA 91125, USA\\
$^{63}$ Dipartimento di Fisica e Astronomia, Universit\'a di Bologna, Via Gobetti 93/2, I-40129 Bologna, Italy\\
$^{64}$ European Space Agency/ESRIN, Largo Galileo Galilei 1, 00044 Frascati, Roma, Italy\\
$^{65}$ ESAC/ESA, Camino Bajo del Castillo, s/n., Urb. Villafranca del Castillo, 28692 Villanueva de la Ca\~nada, Madrid, Spain\\
$^{66}$ School of Physics and Astronomy, Cardiff University, The Parade, Cardiff, CF24 3AA, UK\\
$^{67}$ INAF-IASF Bologna, Via Piero Gobetti 101, I-40129 Bologna, Italy\\
$^{68}$ APC, AstroParticule et Cosmologie, Universit\'e Paris Diderot, CNRS/IN2P3, CEA/lrfu, Observatoire de Paris, Sorbonne Paris Cit\'e, 10 rue Alice Domon et L\'eonie Duquet, 75205, Paris Cedex 13, France\\
$^{69}$ Space Science Data Center, Italian Space Agency, via del Politecnico snc, 00133 Roma, Italy\\
}

\date{\today}
\authorrunning{M. Martinelli et al.}
\titlerunning{\Euclid: constraining dark energy coupled to electromagnetism}

 
\abstract{
{In physically realistic scalar-field based dynamical dark energy models (including, e.g., quintessence) one naturally expects the scalar field to couple to the rest of the model's degrees of freedom. In particular, a coupling to the electromagnetic sector leads to a time (redshift) dependence of the fine-structure constant and a violation of the Weak Equivalence Principle.}
{Here we extend the previous \Euclid forecast constraints on dark energy models to this enlarged (but physically more realistic) parameter space, and forecast how well \Euclid, together with high-resolution spectroscopic data and local experiments, can constrain these models.}
Our analysis combines simulated \Euclid data products with astrophysical measurements of the fine-structure constant, $\alpha$, and local experimental constraints, and includes both parametric and non-parametric methods. For the astrophysical measurements of $\alpha$ we consider both the currently available data and a simulated dataset representative of Extremely Large Telescope measurements and expected to be available in the 2030s.
Our parametric analysis shows that in the latter case the inclusion of astrophysical and local data improves the \Euclid dark energy figure of merit by between $8\%$ and $26\%$, depending on the correct fiducial model, with the improvements being larger in the null case where the fiducial coupling to the electromagnetic sector is vanishing. These improvements would be smaller with the current astrophysical data. Moreover, we illustrate how a genetic algorithms based reconstruction provides a null test for the presence of the coupling.
Our results highlight the importance of complementing surveys like \Euclid with external data products, in order to accurately test the wider parameter spaces of physically motivated paradigms.
}

\keywords{Cosmology: observations -- (Cosmology:) cosmological parameters -- Space vehicles: instruments -- Surveys -- Methods: statistical -- Methods: data analysis}

   \maketitle
%

\section{Introduction}\label{sec:intro}

The search for the physical mechanism underlying the observed low-redshift acceleration of the Universe is a pressing objective of contemporary cosmology. A first task in this endeavour is to map the behaviour of the energy density (or its equation of state parameter) of the dark energy component as a function of redshift -- with the simplest case of a cosmological constant corresponding to a constant energy density. Towards this end, simple parametrizations are often used, commensurate with the limited constraining power of currently available data, e.g. for the dark energy equation of state parameter one has a tight constraint on its present value and a reasonable constraint on its rate of change. Still, these parametrizations should be seen as convenient proxies for more physically realistic models, possibly containing a larger number of model parameters. While such wider parameters spaces are not significantly constrained by current data, they can in principle be constrained by future surveys.

\Euclid is a medium-class mission of the European Space Agency due for launch in 2022. Using a visible imager\,\citep{VIS_paper} and a near-infrared spectrophotometric instrument\,\citep{NISP_paper}, it will perform a photometric and spectroscopic galaxy survey over 15,\,000 squared degrees of extra-galactic sky, plus a deeper survey over 40 squared degrees\,\citep{Redbook}. The main goal of \Euclid is to provide measurements of the geometry of the Universe and the growth of structures up to redshift $z\sim 2$, and beyond. \Euclid will provide three primary cosmological probes: weak gravitational lensing, the clustering of galaxies using measurements from the photometric galaxy survey, and the clustering of galaxies of the spectroscopic survey. The latter will enable precise measurements of the baryon acoustic oscillations and redshift-space distortions. Given the high complementarity of these large-scale structures probes we expect very precise constraints from \Euclid observations, not only on the concordance cosmological constant and cold dark matter ($\Lambda$CDM) model, but also on theoretical extensions of it\,\citep[see e.g.][]{ISTFpaper,Tutusaus2020}.

In \citet{ISTFpaper} (hereafter EC20), the constraining power of \Euclid on dark energy models has been estimated using the common CPL parameterization \citep{Chevallier:2000qy,Linder:2002et} as a phenomenological proxy for generic dynamical dark energy models. However, in physically realistic scalar-field based dynamical dark energy models (including, e.g., quintessence) one naturally expects the scalar field to couple to other sectors of the theory, unless unknown symmetries suppress such a coupling. Here we focus on the possible coupling of a dark energy scalar field to the electromagnetic sector, which would lead to a time (redshift) dependence of the fine-structure constant, $\alpha$, a violation of the Einstein Equivalence Principle \citep{Carroll,Dvali,Chiba}, and also a violation of the distance duality relation. Forecast constraints on the latter from \Euclid and contemporary surveys are discussed in \citet{Martinelli:2020hud}.

There are two immediate consequences of this fact. The first is that one should deal with a wider parameter space: the coupling of the scalar field to the electromagnetic sector is a further relevant parameter, all the more so because, as will be seen in what follows, it is degenerate with the parameters describing the dark energy evolution. The second consequence is that \Euclid, at least with its primary probes, is not able to constrain such a coupling, as its observables are not sensitive to the variation of the fine structure constant, and therefore one needs to add astrophysical and local constraints on $\alpha$ and the Einstein Equivalence Principle to the analysis, in order to test this kind of scenario. A recent review of the synergies between these astrophysical and local tests and cosmological observations is given in \citet{ROPP}.

Therefore our analysis in this work, which builds upon previous works by \citet{Calabrese:2013lga}, \citet{Pinho} and \citetalias{ISTFpaper}, has two main goals:
\begin{itemize}
    \item Forecast how well \Euclid (in combination with external data, specifically  high-resolution spectroscopic data and local experimental results) can constrain these models.
    \item Quantify the change in forecast constraints on CPL parameters when the assumption of a vanishing coupling between the dark energy driving scalar field and electromagnetism is removed.
\end{itemize}

The plan of the rest of the paper is as follows. In \Cref{sec:theory} we review theoretical models relating dynamical dark energy and a varying fine-structure constant. In \Cref{sec:data} we describe the currently available data and the forecast future data used in this work, including both \Euclid measurements and astrophysical and local data. In \Cref{sec:likely} and \Cref{sec:GAanalysis} we describe the analysis methods used in this study: a standard likelihood analysis for the CPL parametrization and a model-independent reconstruction using Genetic Algorithms. We present the results obtained with a likelihood analysis in \Cref{sec:likeres} and the results derived with the Genetic Algorithms in \Cref{sec:GA}. We present our discussion and conclusions in \Cref{sec:summary}.

\section{Dynamical dark energy and varying alpha}\label{sec:theory}

Dynamical scalar fields in an effective four-dimensional field theory are naturally expected to couple to the rest of the theory, unless a still unknown symmetry is postulated to suppress these couplings. In particular, these couplings unavoidably exist in string theory \citep{Taylor,Casas1,Casas2}, and their cosmological role is especially interesting in models where such a dilaton-type scalar field is also responsible for the acceleration of the universe \citep{Carroll,Dvali,Chiba,Dilaton}.  In what follows we will assume this coupling does exist for the dynamical degree of freedom responsible for the dark energy. Specifically we will be interested in the coupling between a canonical scalar field, denoted $\phi$, and the electromagnetic sector, which stems from a gauge kinetic function $B_F(\phi)$
\begin{equation}
{\cal L}_{\phi F} = - \frac{1}{4} B_F(\phi) F_{\mu\nu}F^{\mu\nu}\,.
\end{equation}

Since the local behaviour of electromagnetism is extremely well known and any variations of $\alpha$ are constrained to be very small (as further discussed below) one can safely assume this function to be linear,
\begin{equation}\label{eq:linearexp}
B_F(\phi) = 1- \zeta \kappa (\phi-\phi_0)\,,
\end{equation}
(where we have defined $\kappa^2=8\pi G$) since, as has been pointed out in \citet{Dvali}, the absence of such a term would require the presence of a $\phi\to-\phi$ symmetry. Such a symmetry must be broken throughout most of the cosmological evolution, because $\phi$ is a time-dependent field, changing (possibly very slowly) as the universe expands. In other words, the absence of such a term would require fine-tuning. With this definition $\zeta$ is a dimensionless coupling, which will be crucial in our subsequent discussion. As is physically clear, the relevant parameter in the cosmological evolution is the field displacement relative to its present-day value (in particular $\phi_0$ could be freely set to zero).

With these assumptions one can explicitly relate the evolution of $\alpha$ to that of dark energy, as in \citet{Erminia1}, whose derivation we summarize. The evolution of $\alpha$ can be written as
\begin{equation}\label{eq:zetadef}
\frac{\Delta \alpha}{\alpha} \equiv \frac{\alpha-\alpha_0}{\alpha_0} =B_F^{-1}(\phi)-1=
\zeta \kappa (\phi-\phi_0) \,.
\end{equation}
Defining the fraction of the dark energy density as
\begin{equation}
\Omega_\phi (z) \equiv \frac{\rho_\phi(z)}{\rho_{\rm tot}(z)} \simeq \frac{\rho_\phi(z)}{\rho_\phi(z)+\rho_{\rm m}(z)} \,,
\end{equation}
where in the last step we have neglected the contribution from radiation (since we will be interested in low redshifts, $z<5$, where it is indeed negligible), the evolution of the putative scalar field can be expressed in terms of the dark energy properties $\Omega_\phi$ and $w_\phi$ as \citep{Nunes}
\begin{equation}
1+w_\phi = \frac{(\kappa\phi')^2}{3 \Omega_\phi} \,,
\end{equation}
with the prime denoting the derivative with respect to the logarithm of the scale factor. We finally obtain 
\begin{equation} \label{eq:dalfa}
\frac{\Delta\alpha}{\alpha}(z) =\zeta \int_0^{z}\sqrt{3\Omega_\phi(z')\left[1+w_\phi(z')\right]}\frac{{\rm d}z'}{1+z'}\,.
\end{equation}
The above relation assumes a canonical scalar field, but the argument can be repeated for phantom fields, as discussed in \citet{Phantom}, leading to 
\begin{equation} \label{eq:dalfa2}
\frac{\Delta\alpha}{\alpha}(z) =-\zeta \int_0^{z}\sqrt{3\Omega_\phi(z')\left|1+w_\phi(z')\right|}\frac{{\rm d}z'}{1+z'}\,.
\end{equation}
Physically, the change of sign stems from the fact that one expects phantom fields to roll up the potential rather than down. Naturally, the two definitions match across the phantom divide ($w=-1$), and together they are fully applicable to the CPL parameterization \citep{Chevallier:2000qy,Linder:2002et}. In this case the dark energy equation of state has the form
\begin{equation} \label{cpl}
w_{\rm CPL}(z)=w_0+w_a \frac{z}{1+z}\,,
\end{equation}
while the fraction of energy density provided by the scalar field is easily found to be
\begin{equation}
\Omega_{\rm CPL}(z)=\frac{1-\Omega_{\rm m}}{1-\Omega_{\rm m}+\Omega_{\rm m}(1+z)^{-3(w_0+w_a)}{\rm e}^{3w_az/(1+z)}}\,,
\end{equation}
where we assumed a flat Universe, with a vanishing curvature parameter $\Omega_{\rm K}=0$. When going beyond background probes, the CPL parameterization is commonly combined with the Parametrized Post-Friedmann (PPF) framework for DE perturbations \citep{Hu:2007pj,Hu:2008zd,Fang:2008sn}. The dependence of \Cref{eq:dalfa} and \Cref{eq:dalfa2} on $w_\phi(z)$ also makes it clear that we should expect \citep[as discussed in][]{Calabrese:2013lga} degeneracies between the coupling $\zeta$ and the CPL dark energy parameters, $w_0$ and $w_a$, while the correlation with the matter density should be much weaker. On the other hand, note that the $\alpha$ variation is independent of the Hubble constant.

A varying $\alpha$ violates the Einstein equivalence principle since it clearly violates local position invariance. The realization that varying fundamental couplings also induce violations of the universality of free fall goes back at least to the work of Dicke \citep{Dicke} -- we refer the reader to \citet{Damour} for a recent thorough discussion. The key point in our present context is that a light scalar field, such as the one we are considering here, inevitably couples to nucleons due to the $\alpha$ dependence of their masses, and therefore it mediates an isotope-dependent long-range force. This can be simply quantified through the dimensionless E\"{o}tv\"{o}s parameter $\eta$, which describes the level of violation of the Weak Equivalence Principle (WEP). One can show that for the class of models we are considering here, the E\"{o}tv\"{o}s parameter and the dimensionless coupling $\zeta$ are simply related by \citep{Dvali,Chiba,Damour}
\begin{equation} \label{eq:eotvos}
\eta \approx 10^{-3}\zeta^2\,;
\end{equation}
therefore local experimental constraints on the former can be used to constrain the latter.

We note that there is in principle an additional source term driving the evolution of the scalar field, due to a $F^2B_F'$ term. By comparison to the standard (kinetic and potential energy) terms, the contribution of this term is subdominant, both because its average is zero for a radiation fluid and because the corresponding term for the baryonic density is constrained for the reasons discussed in the previous paragraph. For these reasons, in what follows we neglect this term, which would lead to spatial/environmental dependencies. We nevertheless note that this term can play a role in cosmological scenarios where the dominant standard term is suppressed, such as the models studied in \citet{Olive,Symmetron,Pinho:2017jpk}. 

Finally, another important observable is the current drift rate of the value of $\alpha$, which can easily be found to be
\begin{equation}
\label{eq: drift}
D\equiv\left(\frac{\dot\alpha}{\alpha}\right)_0 = \mp\zeta H_0\sqrt{3 \Omega_{\phi 0} |1+w_0|}\,,
\end{equation}
with the minus and plus signs corresponding respectively to the canonical and phantom cases. Naturally, the drift rate depends on the present value of the dark energy equation of state (and vanishes for $w_0=-1$), but it is independent of $w_a$. This observable provides a second way to constrain these models using local experiments, since the drift rate can be constrained using laboratory experiments which compare atomic clocks based on transitions with different sensitivities to $\alpha$.

\section{Available and future data}\label{sec:data}

The purpose of this work is to constrain canonical scalar-field based dynamical dark energy models which allow for the possible variation of the fine structure constant, as detailed in \Cref{sec:theory}, using both currently available data and the ones expected from future surveys. In particular, we will use data from observations of quasi-stellar object  (QSO) spectral lines from archival datasets and dedicated measurements, complemented by laboratory constraints, detailed in \Cref{sec:curralpha}, as well as forecast data for the future measurements of $\Delta\alpha/\alpha$ from the Extremely Large Telescope (ELT) as discussed in \citet{Leite}; the assumptions made to generate mock datasets for this experiment are shown in \Cref{sec:forealpha}. 

However, as pointed out in \citet{Calabrese:2013lga} and as can also be seen in \Cref{eq:dalfa,eq:dalfa2}, the coupling parameter that drives the variation of $\alpha$ is significantly degenerate with other standard cosmological parameters, namely $\Omega_{\rm m}$, $w_0$ and $w_a$. The constraining power of \Euclid on these parameters is therefore crucial if one wants to constrain this kind of models.

\subsection{Currently available data for $\alpha$ variation}\label{sec:curralpha}

Our astrophysical data consists of high-resolution spectroscopy tests of the stability of $\alpha$. These measurements are done in low-density absorption clouds along the line of sight of bright quasars, typically with wavelength resolution $R=\lambda/\Delta\lambda\sim50\, 000$ (although the exact value is different for different measurements). We use a total of 319 measurements, of which 293 come from the analysis of archival data by \citet{Webb} and the remaining 26 are more recent dedicated measurements \citep{ROPP,Cooksey,Welsh,Milakovic}. The latter subset is therefore smaller than the former, but it contains more stringent measurements, so overall the archival and dedicated subsets have comparable constraining power \citep{Meritxell}. Overall, this dataset includes measurements up to redshift $z\sim4.18$.

Additionally, the current drift rate  of $\alpha$ is constrained by local comparison experiments between atomic clocks, with the most stringent bound being the one by \citet{Lange} 
\be
D\rvert_{\rm obs}\equiv\left(\frac{\dot\alpha}{\alpha}\right)_0=(1.0\pm1.1)\times10^{-18}\,\text{yr}^{-1}\,.
\ee

Last but not least, we also use the recent MICROSCOPE bound on the E\"otv\"os parameter of \citet{Touboul}
\be
\eta=(-0.1\pm1.3)\times10^{-14}\,,
\ee
which, as previously discussed, constrains the model's coupling to the electromagnetic sector.

In the rest of the paper, we refer to the combination of all these data as {\it current $\alpha$ data}, and we will show the constraints produced by such a combination. However, we note that the 293 archival data an the 26 dedicated ones are in slight tension with each other \citep{ROPP,Meritxell}. Despite assuming here that they can be safely combined, we discuss this issue in more detail in \Cref{sec:app_current}.

In our analysis, we do not include geophysical constraints on the variation of $\alpha$, coming from the Oklo natural nuclear reactor \citep{Fujii:2002hc,Davis:2015ila} and meteoric data \citep{Olive:2003sq}. This is motivated by the model dependence of such data, as they only provide stringent constraints on $\alpha$ if one assumes that only the fine-structure constant can vary while the strong sector of the theory is unchanged, which is a simplistic assumption for constraints that stem from nuclear-physics processes \citep{ROPP}. Thus, they are less reliable than the spectroscopic and atomic clock data.

\subsection{ELT forecast}\label{sec:forealpha}

Here we assume a future dataset to be put together by the high-resolution ultra-stable spectrograph currently known as HIRES \citep{SPIE}, that will operate the 39.3 meter Extremely Large Telescope. For simplicity we assume a set of 50 $\alpha$ measurements, uniformly spaced in the redshift range $0.7\leq z\leq3.2$, each with an uncertainty of 0.03 parts per million, which is commensurate with the assumptions in \citet{Leite} and the Top-Level Requirements for the instrument \citep{HIRES}.

With these specifications in hand, we generate the fiducial redshift dependence of $\Delta\alpha/\alpha$ using \Cref{eq:dalfa,eq:dalfa2} with two different fiducial cosmologies, dubbed $\Lambda$CDM and $\zeta w_0w_a$CDM, shown in \Cref{tab:fiducial}. These two fiducial cosmologies correspond to a standard case in which no $\alpha$ variation is present ($\Lambda$CDM) and to one where instead we assume the coupling parameter $\zeta$ is non vanishing, but still compatible with the laboratory constraints discussed in \Cref{sec:curralpha}, and a dark energy component that does not behave as a cosmological constant. In this second case ($\zeta w_0w_a$CDM) the value of $\alpha$ varies in redshift and we aim at finding signatures of such variation. 

Once the fiducial behaviour for $\Delta\alpha/\alpha$ is obtained as discussed earlier, we then create the corresponding mock dataset drawing the data points at each redshift from a Gaussian distribution centred at the fiducial model and with $\sigma$ the expected observational error of HIRES. We assume such errors to be uncorrelated, which observationally is a safe assumption since each measurement comes from a high-resolution ($R\sim 100\,000$) signal-to-noise limited spectrum of a point source along a different line of sight. 

\begin{table}
\centering
\caption{Fiducial values for the two cosmologies considered here and used to obtain the mock datasets for ELT measurements. Here $h$ is the reduced Hubble parameter, corresponding to $H_0/(100$ km\,s$^{-1}$Mpc$^{-1}$$)$. \label{tab:Euclidfiducial}}
\begin{tabular}{ccc} 
\hline 
Parameter symbol & $\Lambda$CDM & $\zeta w_0w_a$CDM\\
\hline 
 $\Omega_{\rm m}$ & $0.32$ & $0.32$\\
 $h$ & $0.67$ & $0.67$\\
 $w_0$ & $-1.$ & $-0.94$ \\
 $w_a$ & $0$ & $0.1$ \\
 $\zeta$ & $0$ & $-5\times10^{-8}$ \\
\hline
\end{tabular}
\label{tab:fiducial}
\end{table}

\subsection{\Euclid forecast methodology and Fisher matrices}\label{sec:fisher}


As previously discussed, observations from Large Scale Structures probe are not very sensitive to variations of $\alpha$ and therefore they cannot significantly constrain the coupling $\zeta$. They are however crucial to break the degeneracies between the coupling and the cosmological parameters, and can be therefore combined with the datasets discussed above. In this work we consider specifically \Euclid as our probe of LSS. It is worth mentioning that, strictly speaking, LSS probes can provide some relevant constraints on variations of $\alpha$. In \citet{Albareti2015}, for example, constraints on $\Delta \alpha/\alpha$ were provided using the O{\sc iii} doublet from BOSS DR12 quasar spectra. Following similar approaches, we could extract information on the variation of $\alpha$ from the future \Euclid data. Furthermore, a type-Ia supernovae survey using \Euclid data\,\citep{AstierDESIRE} could provide some information on the variation of $\alpha$, as illustrated in \citet{Calabrese:2013lga}. However, we prefer to focus here on the main \Euclid probes and their constraints on the cosmological parameters.

In this work, in order to forecast the constraints from the future \Euclid data, we follow the methodology presented in \citetalias{ISTFpaper}. We consider a Fisher matrix formalism and make use of the \texttt{TotallySAF}\,\footnote{\url{https://github.com/syahiacherif/TotallySAF_Alpha}} code \citep{YahiaCherif2020,Tutusaus2020} validated therein for the main \Euclid probes: spectroscopic galaxy clustering (GCsp), photometric galaxy clustering (GCph), weak lensing (WL), and the cross-correlation (XC) terms between the photometric probes. As was done in \citetalias{ISTFpaper}, we neglect any correlation between the spectroscopic and photometric probes.

Starting with the spectroscopic probe, we build a Fisher matrix for the observed anisotropic power spectrum of H-$\alpha$ emitters \citepalias[see Eq. 87 in][]{ISTFpaper}, accounting for a phenomenological model for non-linearities, the Alcock-Paczynski effect, redshift-space distortions, and the Fingers-of-God effect. As in \citetalias{ISTFpaper}, we consider two scenarios for these forecasts. In the optimistic case, we consider all scales up to a maximum of $k_{\rm max}=0.30\,h\,\text{Mpc}^{-1}$, and we fix the nuisance parameters associated to non-linearities. In the pessimistic scenario we limit our analysis to scales $k<k_{\rm max}=0.25\,h\,\text{Mpc}^{-1}$ and marginalize over the non-linear nuisance parameters.

With respect to the photometric probes, we build a Fisher matrix for the tomographically binned projected angular power spectra. The same formalism is used for WL, GCph, and their XC terms, with the only difference being the kernels used in the projection from the power spectrum of matter perturbations to the spherical-harmonic space observable. As in \citetalias{ISTFpaper}, we use the Limber, flat-sky and spatially flat approximations \citep{Kitching2017,Kilbinger2017,Taylor2018}. We also neglect redshift-space distortions, magnification, and other relativistic effects \citep{Deshpande2020}, but marginalize over the galaxy bias and intrinsic alignment nuisance parameters. We refer to \citetalias{ISTFpaper} for all the details on the modelling. As in the spectroscopic case, we consider two different scenarios in our forecasts. In the optimistic setting, we consider all multipoles between $\ell_{\rm min}=10$ and $\ell_{\rm max}=3000$ for GCph and the XC terms and $\ell_{\rm max}=5000$ for WL. We then combine with the spectroscopic constraints assuming they are independent. In the pessimistic scenario, we limit the multipoles to $\ell_{\rm max}=750$ for GCph and the XC terms and $\ell_{\rm max}=1500$ for WL. In this case, when combining with the spectroscopic probe, we introduce a redshift cut of $z<0.9$ for GCph and the XC terms, in order to remove any possible correlation with the spectroscopic sample starting at $z=0.9$.

In accordance with \Cref{sec:theory} and \citetalias{ISTFpaper}, we consider in this work a cosmological model with a dark energy equation of state parametrized with the CPL parametrization and with $\Omega_{\rm K}=0$. In contrast with \citetalias{ISTFpaper}, here we do not consider just a single $\Lambda$CDM fiducial (since under the assumptions of the present work there would be no $\alpha$ variation in that case), and obtain the Fisher matrices in both assumed cosmologies of \Cref{tab:fiducial}. In addition to these assumed parameters, the analysis of \citetalias{ISTFpaper} also requires to specify the fiducial values of other cosmological parameters, namely the baryon energy density ($\Omega_{\rm b}=0.05$), the primordial spectral index ($n_{\rm s}=0.96$) and the current amplitude of density perturbations ($\sigma_8=0.816$). These additional parameters take the same fiducial values in both the $\Lambda$CDM and $\zeta w_0w_a$CDM cosmologies.

For completeness, and in order to compare with the results of this work, we provide in \Cref{tab:Euclid_baseline} the baseline \Euclid forecasts obtained in \citetalias{ISTFpaper} for the relevant parameters in our analysis. We also note that the current constraints on the dark energy equation of state parameters are $w_0=-0.957\pm 0.080$ and $w_a=-0.29^{+0.32}_{-0.26}$, using the combination of Planck 2018 temperature, polarization, and lensing measurements, together with type-Ia supernovae and baryon acoustic oscillations observations\,\citep{Planck2018}.

\begin{table}
\centering
\caption{Baseline \Euclid forecast uncertainties for the relevant parameters in this work, $\Omega_{\rm m}$, $h$, $w_0$, and $w_a$, obtained in \citetalias{ISTFpaper}.}
\begin{tabular}{ccc} 
\hline 
Parameter symbol & Pessimistic & Optimistic\\
\hline 
 $\sigma(\Omega_{\rm m})$ & $0.0038$ & $0.0018$\\
 $\sigma(h)$ & $0.0037$ & $0.0010$\\
 $\sigma(w_0)$ & $0.040$ & $0.025$ \\
 $\sigma(w_a)$ & $0.17$ & $0.092$ \\
\hline
\end{tabular}
\label{tab:Euclid_baseline}
\end{table}

\section{Likelihood analysis for the CPL parametrization}\label{sec:likely}

Following the prescription for a possible $\alpha$ variation described in \Cref{sec:theory}, we want to combine current and forecast $\alpha$ measurements with the information that will be brought by \Euclid, thus investigating how this survey will improve our constraints on this possible deviation from the standard cosmological paradigm.

While in \Cref{sec:fisher} we discussed how \Euclid constraints can be predicted using the Fisher matrix approach, the strong non-Gaussian nature of the joint cosmology and fundamental physics parameter space in varying $\alpha$ models \citep[][]{Calabrese:2013lga} makes this approach unfeasible for both the current and future measurements that we are interested in.

Therefore, we rely here on an MCMC approach, using the publicly available sampler \texttt{Cobaya} \citep{Torrado:2020dgo}, which exploits a Metropolis-Hastings (MH) algorithm \citep{Lewis:2002ah,Lewis:2013hha}. We sample the matter density parameter $\Omega_{\rm m}$, the Hubble constant $H_0$, the two parameters of the CPL parameterization $w_0$ and $w_a$, and the coupling parameter $\zeta$ that connects the dynamical dark energy scalar field to the electromagnetic sector (see \Cref{sec:theory}). The posterior distribution $P(\theta)$ that we reconstruct with this method at each point $\theta=(\Omega_{\rm m},H_0,w_0,w_a,\zeta)$ of the parameter space contains information coming from both $\alpha$ measurements and the \Euclid survey,
\begin{equation}\label{eq:posterior}
    P(\theta) \propto \mathcal{L}_\alpha(\theta)\mathcal{L}_\Euclid(\theta)\, ,
\end{equation}
where $\mathcal{L}_\alpha$ and $\mathcal{L}_\Euclid$ are the likelihoods of the $\alpha$ and \Euclid datasets respectively, and we assumed that the two probes are uncorrelated.

The \Euclid likelihood is constructed using the Fisher matrix $\tens{F}$ described in \Cref{sec:fisher}, and it simply exploits the Gaussian assumption done to obtain these:
\begin{equation}\label{eq:euclike}
    -\ln{\mathcal{L}_\Euclid}\propto\frac{1}{2}(\theta-\theta_{\rm fid})^{\rm T}\tens{\tilde{F}}(\theta-\theta_{\rm fid}),
\end{equation}
where $\tens{\tilde{F}}$ is the Fisher matrix $\tens{F}$ marginalized over all parameters that are not contained in our sampled parameter space, while $\theta_{\rm fid}$ is the fiducial cosmology under examination, which can be one of the two shown in \Cref{tab:fiducial}.

On the other hand, the $\alpha$ likelihood contains two different contributions, again assumed to be uncorrelated, with
\begin{equation}\label{eq:alphalike}
    -\ln{\mathcal{L}_\alpha} \propto -\left(\ln{\mathcal{L}_{\rm QSO}}+\ln{\mathcal{L}_{\rm clocks}}\right)\, ,
\end{equation}
with the first contribution given by observations of quasar absorption systems and the second, coming from atomic clocks measurements, giving a constraint on the possible coupling $\zeta$ at present time. 
These two likelihoods are taken to be
\begin{equation}\label{eq:likeqso}
    -\ln{\mathcal{L}_{\rm QSO}} \propto \frac{1}{2}\sum_i{\frac{1}{\sigma_i^2}\left[\frac{\Delta\alpha}{\alpha}\bigg\rvert_{\rm th}(z_i)-\frac{\Delta\alpha}{\alpha}\bigg\rvert_{\rm obs}(z_i)\right]^2}\,,
\end{equation}
and
\begin{equation}\label{eq:likeclocks}
    -\ln{\mathcal{L}_{\rm clocks}} \propto \frac{1}{2}\frac{\left(D\rvert_{\rm th}-D\rvert_{\rm obs}\right)^2}{\sigma_{\rm D}^2}\,,
\end{equation}
where the ${\rm th}$ subscript indicates the theoretical predictions, given by \Cref{eq:dalfa,eq:dalfa2} for $\Delta\alpha/\alpha$ and \Cref{eq: drift} for $D$, while the quantities labelled with the ${\rm obs}$ subscript and the corresponding errors are the measurements described in \Cref{sec:data}.

We therefore sample the parameter space described above and reconstruct the posterior of \Cref{eq:posterior}, using flat priors on all parameters except for the coupling $\zeta$, for which a Gaussian prior centred in $\zeta=0$ and with variance $1.3\times10^{-14}$ is used. Such prior information is derived from the MICROSCOPE experiment discussed in \Cref{sec:curralpha}, which directly constrains the E\"otv\"os parameter $\eta$, which is related to the coupling $\zeta$ via \Cref{eq:eotvos}.

In addition, we obtain as derived parameters also the value of $\Delta\alpha/\alpha$ in a set of equally spaced redshifts $z_i$. Obtaining the marginalized mean values of these derived parameters and their $68\%$ confidence limit, we will reconstruct the trend of the variation of $\alpha$ with redshift in \Cref{sec:likeres}.

\section{Genetic Algorithm analysis}\label{sec:GAanalysis}

In our analysis we also use a non-parametric machine learning class of stochastic optimization methods, known as Genetic Algorithms (GA). These emulate natural selection, by using the data as proxies for the evolutionary pressure that drives the selection of the best-fitting functions in each generation. They are characterized by the notion of grammatical evolution, as described by the genetic operations of mutation and crossover. Specifically, a set of functions will evolve over time under the pressure of the data and the influence of the stochastic operators of crossover, i.e. the combination of different functions to form more complicated forms (offspring) that may fit the data better, and mutation, namely a random change in an individual function.

The GA have been used extensively to test for extensions of the standard model \citep[]{Akrami:2009hp}, deviations from the cosmological constant model, both at the background and the perturbations level \citep[]{Nesseris:2012tt,Arjona:2019fwb,Arjona:2020kco}, to reconstruct a plethora of cosmological data, such as type Ia supernovae or CMB  \cite[]{Bogdanos:2009ib, Arjona:2020doi} or to reconstruct various null tests such as the so called Om statistic or the curvature test \citep[]{Nesseris:2010ep, Nesseris:2013bia, Sapone:2014nna}.

In our analysis we fit the fine-structure data with the GA, however we solely focus on the coupling $\zeta$. From \Cref{eq:zetadef} it is clear that $\zeta$ should be a constant within the context of the non-minimal coupling to the Maxwell field and the linear approximation of the gauge kinetic function $B_F(\phi)$, so we use the GA to test whether this assumption is actually supported by the data. In other words, we will treat the constant $\zeta$ case as a null test and use the GA to test for deviations from that behaviour. This approach has the advantage that the coupling $\zeta$ is directly related to measurable quantities, especially in the case of the local experiments, see for example \Cref{eq:eotvos}.

In detail, the analysis of the data with the GA proceeds as follows. First, we assume that the probability that a given function in the population will produce offspring, or equivalently its “reproductive success”, is proportional to its fitness. We quantify this fitness via a $\chi^2$ statistic, which is obtained following the same likelihood computation used in \Cref{sec:likely}. Second, an initial random group of functions is chosen, each representing an initial guess for the coupling, though they are allowed to be redshift dependent, i.e. $\zeta=\zeta(z)$. 

We note that as the $\alpha$ data extends to high redshifts, we base our GA grammar not in terms of the redshift $z$, but instead in terms of $1-a=\frac{z}{1+z}$, something which is commonly used in other model independent methods as well, see for example  \citet{Cattoen:2007id,Lazkoz:2013ija,Guimaraes:2010mw}. This allows us to avoid any spurious reconstructions due to lack of convergence at high redshifts.

Then, in the case of the current data, $\Delta \alpha/\alpha$ can be related directly to $\zeta$ and compared to the data using \Cref{eq:dalfa,eq:dalfa2}, for which we need to estimate the integral in the right hand side which can be done with the information provided by \Euclid. Note that this integral is by default zero when the fiducial model is exactly the cosmological constant $\Lambda$CDM model, which means that the best-fit $\zeta$ will remain indeterminate, so in our analysis we only use the \Euclid $\zeta w_0w_a$CDM fiducial model, as given in in \Cref{tab:Euclidfiducial}. We then use 300 realizations and we calculate both the mean and variance of the integral, with the latter then included, via error propagation, in the error estimate of all reconstructed quantities. For the atomic clocks and the MICROSCOPE bound we use \Cref{eq:eotvos}, \Cref{eq: drift} and the \Euclid $\zeta w_0w_a$CDM fiducial in a similar fashion. 

After this is done, we can compare the predictions of the GA with the data and the fitness of every test function in the population can be calculated via a standard $\chi^2$ statistic. Subsequently, the crossover and mutation operators are applied to a subset, usually the $\sim 30 \%$ best-fitting functions in every generation. These are chosen with the tournament selection -- see \citet{Bogdanos:2009ib} for more details. We repeat this process thousands of times in order to ensure convergence and we also test our fits with several different random seeds, so as not to bias the results.

As soon as the GA has converged, the given best-fit function is an analytic and smooth function of the redshift $z$ that describes the possible evolution of the coupling $\zeta(z)$. 

The errors on this best-fit are estimated using an analytical approach developed by \citet{Nesseris:2012tt,Nesseris:2013bia}, in which the errors are estimated by a path integral over the whole functional space. This approach has been exhaustively tested by \citet{Nesseris:2012tt} and has been found to be in very good agreement with Monte Carlo based error estimates.

The fact that in this approach we allow for a redshift dependence of the coupling $\zeta$ allows us to examine whether our assumptions, which rely on a constant coupling, are still valid or they break down. This is done with the goal of obtaining a null test for the constancy of $\zeta$ and the linear expansion of the gauge kinetic function in \Cref{sec:theory}. Indeed, should a variation of $\alpha$ be supported by the data, but generated by a mechanism that violates our assumptions, \Cref{eq:dalfa,eq:dalfa2} would not be able to model its redshift trend, and our reconstruction would yield a non-constant coupling $\zeta(z)$. Any statistically significant deviations from a constant value at any redshift will imply that our original hypothesis of a constant coupling may be violated. This is analogous (though not identical) to the reconstruction of a parameter quantifying possible distance duality violations, discussed in \citet{Martinelli:2020hud}. Effectively, our approach will allow us to determine whether or not the data is consistent with  a null value of the coupling, and how well this is constrained at various redshifts. This is also akin to constraining the behaviour of the coupling in different redshift bins.

In this work, the specific numerical implementation of the GA is based on the publicly available code \texttt{Genetic Algorithms}\footnote{\url{https://github.com/snesseris/Genetic-Algorithms}}. For more details and how they apply to the analysis of \Euclid data, see also \citet{Martinelli:2020hud}.

\section{Likelihood approach results}\label{sec:likeres}

In this Section, we show the results of the analysis presented in \Cref{sec:likely}. We recall that our goal is to illustrate how the use of external data allows \Euclid to constrain dynamical dark energy models including an electromagnetic sector coupling. Specifically, the external data will constrain the coupling $\zeta$, to which \Euclid itself is insensitive. We first discuss the result of combining \Euclid with currently available $\Delta\alpha/\alpha$ and other current data, and focus on the analogous results when \Euclid data is combined with next generation high-resolution spectroscopy data, specifically that expected from the ELT. Finally, we show the results of a Bayesian evidence analysis that quantifies the possible significance of a detection of a varying fine-structure constant.

\subsection{\Euclid and current $\alpha$ measurements}\label{sec:likecurr}

\begin{figure}
    \centering
    \includegraphics[width=0.9\columnwidth]{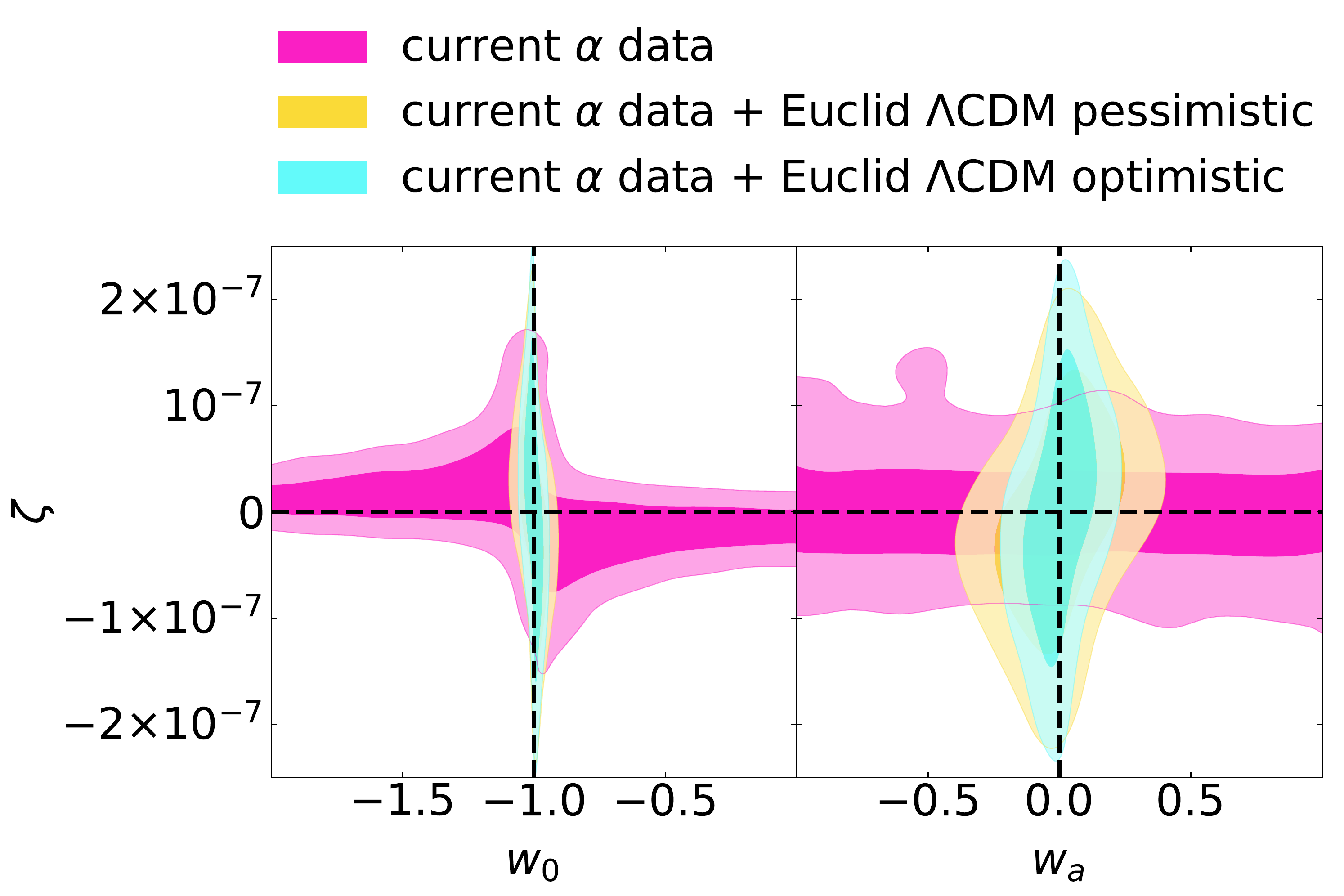}\\
    \includegraphics[width=0.9\columnwidth]{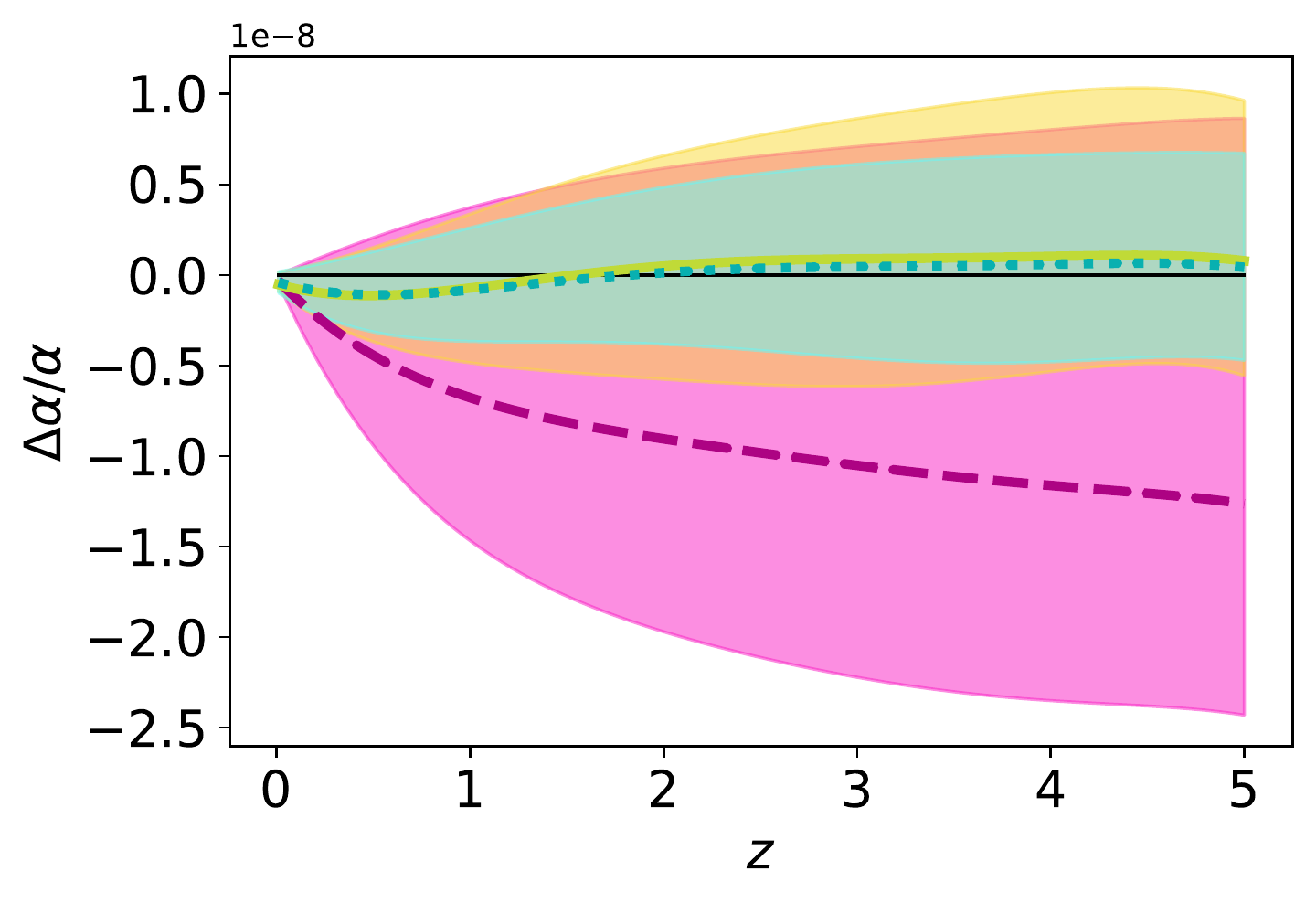}
    \caption{{\it Top panel}: constraints on the CPL and coupling parameters using currently available data for $\alpha$ measurements alone (purple contours) and in combination with \Euclid forecast constraints with a $\Lambda$CDM fiducial (yellow contours for the pessimistic case and cyan contours for the optimistic case). {\it Bottom panel}: reconstruction of the mean trend in redshift of $\Delta\alpha/\alpha$ and of the allowed $68\%$ confidence level area, obtained interpolating the marginalized means and errors of the derived parameters described in \Cref{sec:likely}. The purple dashed line and purple area refer to $\alpha$ measurements alone, the yellow solid line and yellow area include \Euclid in the pessimistic case, while the dotted green line and green contours combine the optimistic case. }
    \label{fig:WebbLCDM}
\end{figure}

\cref{fig:WebbLCDM} shows in purple the model parameter constraints obtained using currently available $\alpha$ measurements, comprising the combination of Webb archival data and the dedicated $\alpha$ measurements, atomic clocks constraints and the MICROSCOPE bound. The top panel shows the constraints on the free CPL and coupling parameters, while the bottom one highlights the reconstruction of the redshift trend of $\Delta\alpha/\alpha$. We show the results without and with \Euclid data, for a $\Lambda$CDM fiducial, and for the \Euclid data we show in yellow the pessimistic case and in cyan the optimistic one. 

In \Cref{tab:currres}, it is possible to notice how the constraint on the coupling $\zeta$ becomes less stringent when \Euclid is included, changing from $\zeta=\left(\,0.1^{+3.5}_{-3.9}\,\right)\times10^{-8}$ to $\zeta=\left(\,-0.1\pm 8.2\,\right)\times10^{-8}$ in the pessimistic case, and $\zeta=\left(\,0.4\pm 8.9\,\right)\times10^{-8}$ when the \Euclid optimistic configuration is used. The reason for such loosening of the bound is due to the ability of \Euclid to tightly constrain the CPL parameters around the $\Lambda$CDM limit $(w_0,w_a)=(-1,0)$; due to the degeneracy with $\zeta$ -- see \Cref{eq:dalfa} and \Cref{eq:dalfa2} -- this makes the $\alpha$ data less sensitive to the coupling parameter, as now a wider range of values is able to fit the data. Such a result is compatible with what was found in \citet{Calabrese:2013lga}, where other datasets able to tightly constrain the CPL parameters were considered. In the bottom panel of \Cref{fig:WebbLCDM}, it can be seen however how the weaker constrain on $\zeta$ does not lead to a larger spread of the allowed $\Delta\alpha/\alpha$ reconstructions, which are instead tightly constrained around the no-variation limit by the inclusion of \Euclid information, exactly because of the tight constraints on the CPL parameters.

\begin{figure}
    \centering
    \includegraphics[width=0.9\columnwidth]{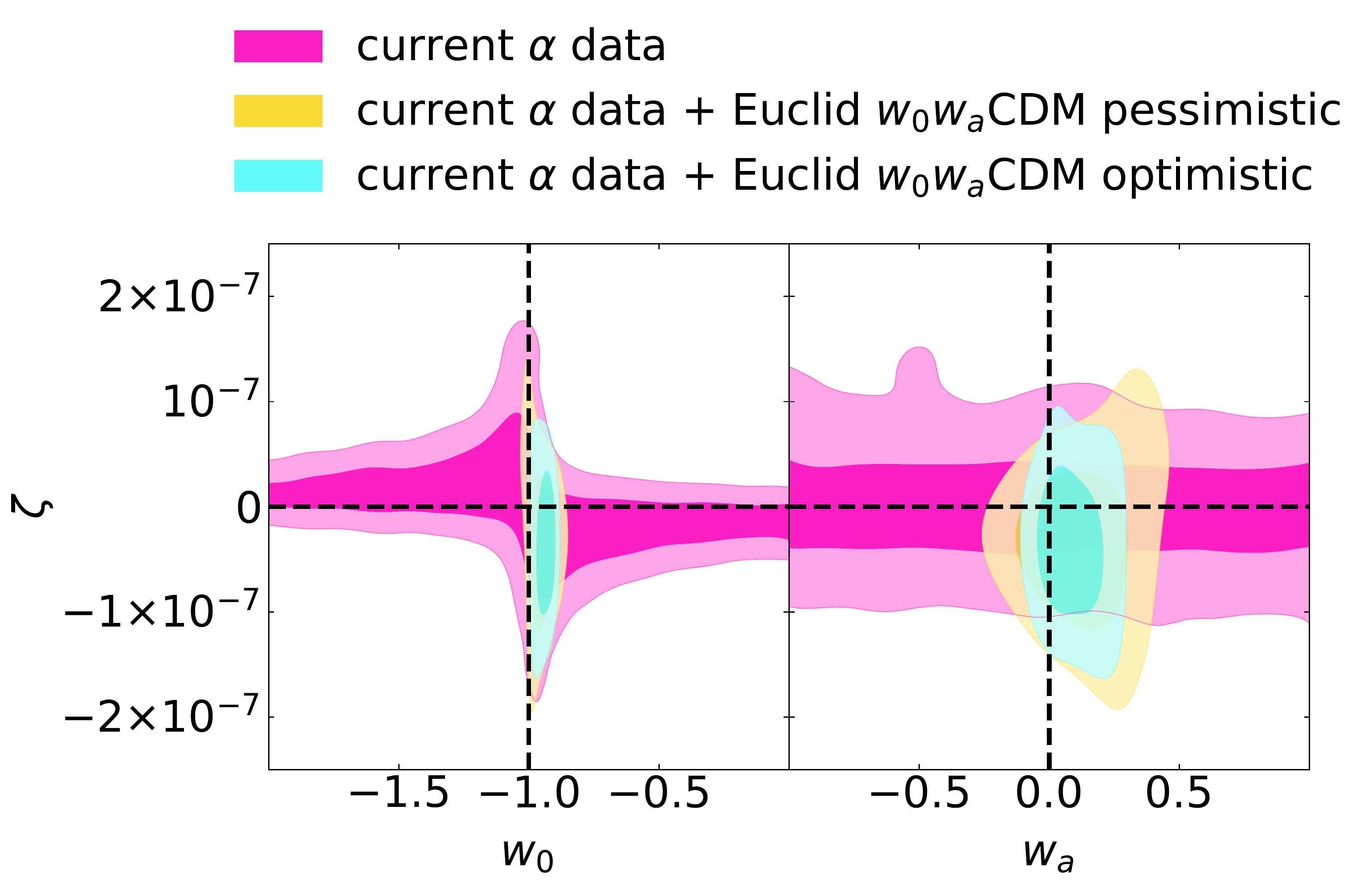}\\
    \includegraphics[width=0.9\columnwidth]{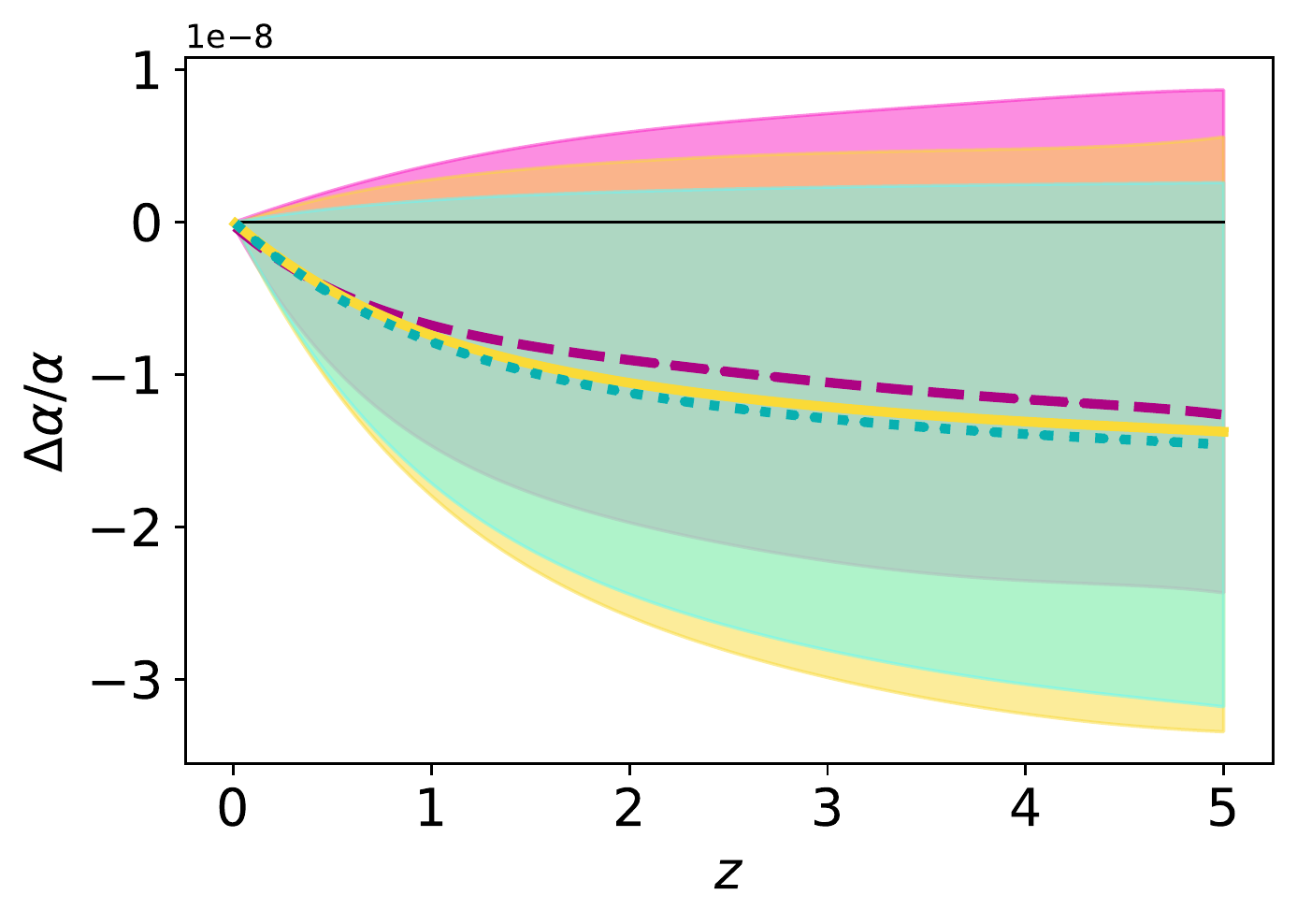}
    \caption{Same as \Cref{fig:WebbLCDM}, but when the fiducial used to obtain the \Euclid results is $\zeta w_0w_a$CDM. Notice that here we use currently available direct measurements of $\alpha$, which are not impacted by our choice of fiducial cosmology. Thus, the modified $\zeta$ fiducial value does not affect these results, as the \Euclid probes considered are not sensitive to this parameter.}
    \label{fig:WebbDE}
\end{figure}

The situation changes when the \Euclid results obtained for the $\zeta w_0w_a$CDM fiducial cosmology are used. In \Cref{fig:WebbDE} it is possible to notice how in this case the \Euclid bound centred on a non-$\Lambda$CDM value of the CPL parameters breaks the degeneracy between these and $\zeta$. Here the loosening of the constraint on the coupling parameter is reduced, with the bound changing from $\zeta=\left(\,0.1^{+3.5}_{-3.9}\,\right)\times 10^{-8}$ to $\zeta=\left(\,-3.6\pm 5.9\,\right)\times 10^{-8}$ (pessimistic) and $\zeta=\left(\,-3.8\pm 4.9\,\right)\times 10^{-8}$ (optimistic), and the constraint on the redshift trend of $\Delta\alpha/\alpha$ is tightened around a non-vanishing variation of the fine-structure constant. The different behaviour of the bounds on $\zeta$ when \Euclid is included in the analysis is due here to the fact that the LSS information constrain the $w_0$ and $w_a$ parameters away from the $\Lambda$CDM limit; this implies that the degeneracy shown in the previous case is broken, and the $\alpha$ data do not have a larger range of allowed coupling value.

While one can see that the inclusion of \Euclid data will help $\alpha$ measurements constrain the variation of this fundamental parameter (since it provides information on the cosmological parameters), the synergy between these two datasets goes both ways: given the assumption done in \Cref{sec:theory} that the scalar field responsible for dark energy is the one that couples with the electromagnetic sector, the $\alpha$ measurements also improve \Euclid constraints on the CPL parameters, since \Euclid on its own is not sensitive to $\zeta$. While this improvement is not extreme, it can be noticed in our results (see \Cref{tab:currres}), where the errors on $w_0$ and $w_a$ are slightly improved with respect to the \citetalias{ISTFpaper} bounds shown in \Cref{sec:fisher}. The additional constraining power leads to an increase of the Figure of Merit (FoM), which we define as \citepalias{ISTFpaper}
\begin{equation}
    {\rm FoM} = \sqrt{\det{(C_{w_0,w_a})^{-1}}}\, ,
\end{equation}
with $C_{w_0,w_a}$ the covariance matrix, obtained from our MCMC results, marginalized over all parameters except for the CPL ones. In the $\Lambda$CDM fiducial, the combination of \Euclid and $\alpha$ data improves the FoM by $18\%$ ($13\%$) with respect to the \Euclid pessimistic (optimistic) value alone. When instead the fiducial for $w_0$ and $w_a$ is shifted from the cosmological constant limit, such improvement becomes $3\%$ in both the pessimistic and optimistic cases. Such a result comes from our assumption that the DE field is the one responsible for the variation of $\alpha$; this relation makes the astrophysical data sensitive to the DE parameters $w_0$ and $w_a$, while no contributions to the FoM would be added if the $\alpha$ variation is not related to DE (or if one assumes a fixed vanishing coupling $\zeta$).

\begin{table}
\centering
\caption{Mean values and $68\%$ c.l. bounds obtained using current $\alpha$ measurements and their combination with \Euclid forecasts. Notice that in this case, as current $\alpha$ data are used, the modified $\zeta$ fiducial value of $\zeta w_0w_a$CDM does not affect the results, given that the fiducial cosmology is used only for \Euclid data, which are not sensitive to this parameter.}
\resizebox{\columnwidth}{!}{
\begin{tabular}{cccc} 
\hline
    \multicolumn{4}{c}{$\Lambda$CDM fiducial} \\
                  & current $\alpha$    & +\Euclid pess      &  +\Euclid opt \\
 \hline 
$\Omega_{\rm m}$  & ---                 & $0.3200\pm 0.0035$ & $0.3199\pm 0.0018$\\

$w_0$             & $-0.99\pm 0.48$     & $-1.000\pm 0.036$  & $-1.001\pm 0.023$\\

$w_a$             & ---                 & $0.00\pm 0.15$     & $0.002\pm 0.087$\\

$H_0$             & $< 75.2$            & $67.00\pm 0.36$    & $67.00\pm 0.10$\\

$\zeta\,10^8$     & $0.1^{+3.5}_{-3.9}$ & $-0.1\pm 8.2$      & $0.4\pm 8.9$\\
\hline
     \multicolumn{4}{c}{$\zeta w_0w_a$CDM fiducial} \\
                 & current $\alpha$    & +\Euclid pess              &  +\Euclid opt \\
 \hline 
$\Omega_{\rm m}$ & ---                 & $0.3194\pm 0.0038$         & $0.3197\pm 0.0019$\\

$w_0$            & $-0.99\pm 0.48$     & $-0.948^{+0.034}_{-0.041}$ & $-0.944\pm 0.024$\\

$w_a$            & ---                 & $0.13^{+0.15}_{-0.14}$     & $0.112\pm 0.085$\\

$H_0$            & $< 75.2$            & $66.98\pm 0.37$            & $67.00\pm 0.10$\\

$\zeta\,10^8$    & $0.1^{+3.5}_{-3.9}$ & $-3.6\pm 5.9$              & $-3.8\pm 4.9$\\
\hline
\end{tabular} 
}
\label{tab:currres}
\end{table}

\subsection{\Euclid and the ELT}

After quantifying the impact of current constraints on $\alpha$ on \Euclid, we now focus on the synergy between \Euclid and the next generation high-resolution spectrograph for the ELT. 

\begin{figure}
    \centering
    \includegraphics[width=0.9\columnwidth]{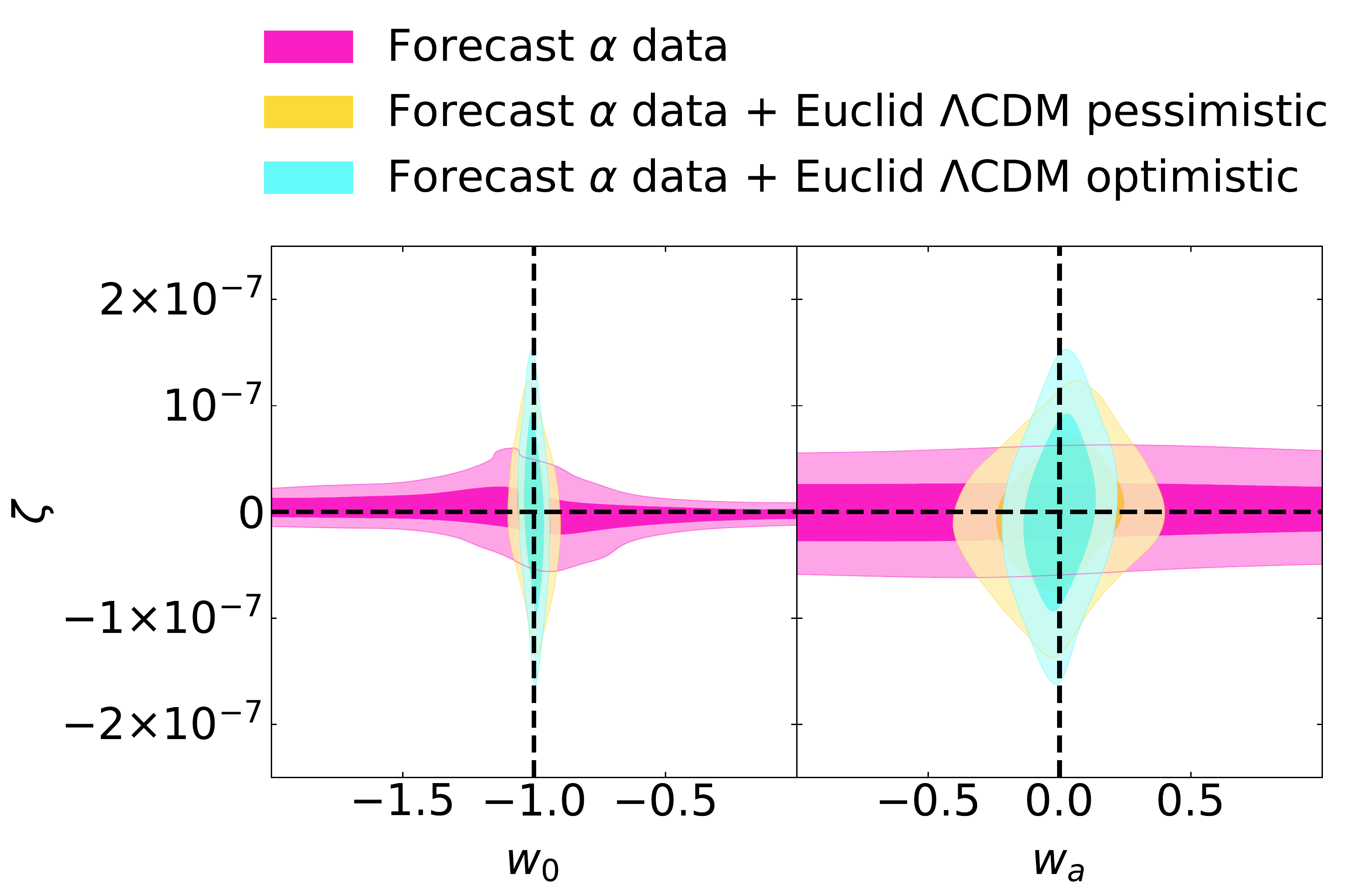}\\
    \includegraphics[width=0.9\columnwidth]{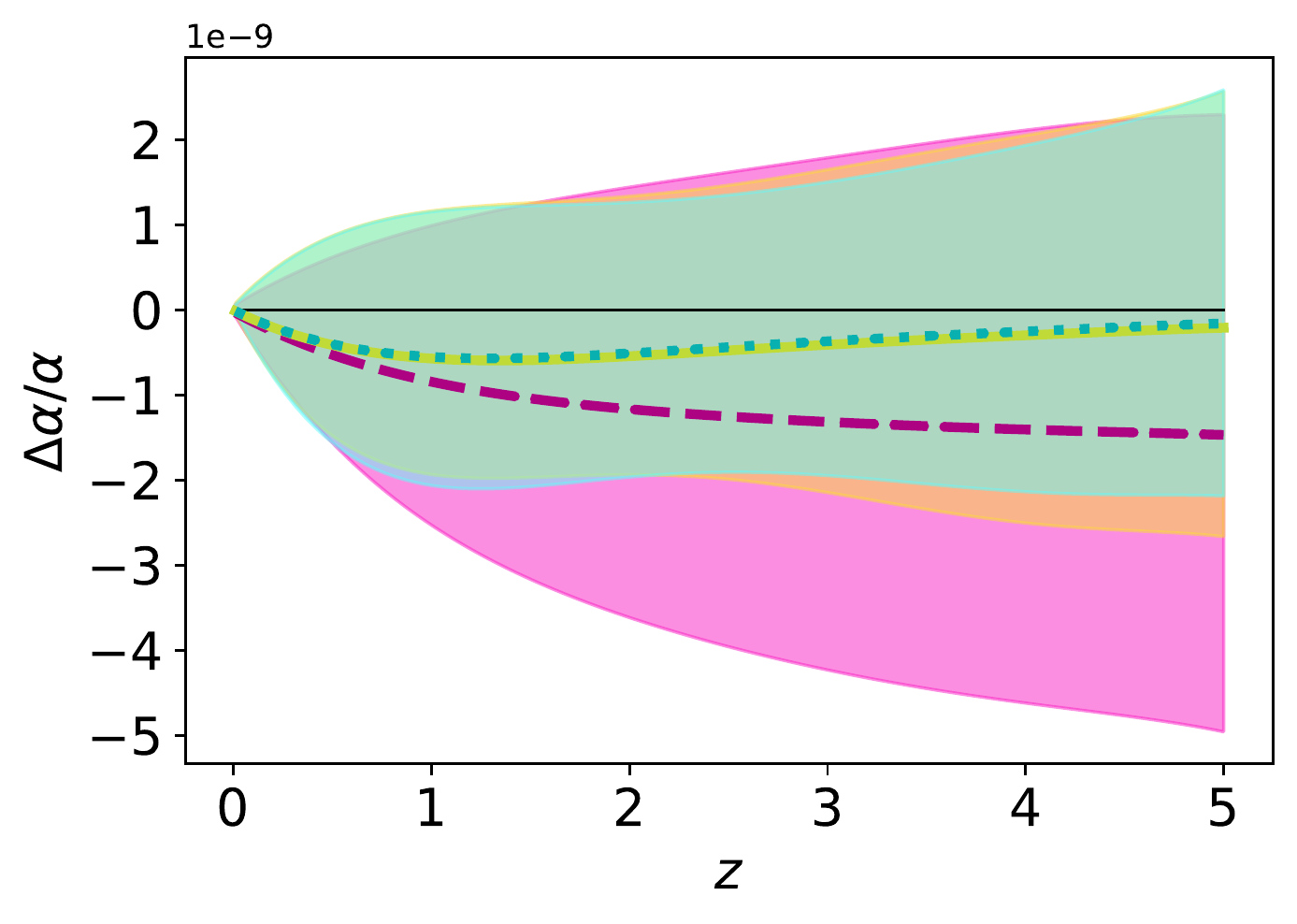}
    \caption{Same as \Cref{fig:WebbLCDM}, but here the purple contours and lines refer to the forecast ELT data, while the yellow and cyan refer to the combination of this simulated dataset with pessimistic and optimistic \Euclid results. The fiducial used here for both ELT and \Euclid is the $\Lambda$CDM one shown in \Cref{tab:fiducial}.}
    \label{fig:ELTLCDM}
\end{figure}

In \Cref{fig:ELTLCDM} we show the results for the $\Lambda$CDM fiducial, with ELT constraints in purple, and those with the inclusion of pessimistic and optimistic \Euclid data in yellow and cyan respectively. As for the current data case, we find that the inclusion of \Euclid leads to a loosening of the constraints on the coupling parameter, see \Cref{tab:forres}, but with a tightening of the reconstruction of $\Delta\alpha/\alpha$ around the $\Lambda$CDM limit due to the information on $w_0$ and $w_a$ brought by \Euclid. 

\begin{figure}
    \centering
    \includegraphics[width=0.9\columnwidth]{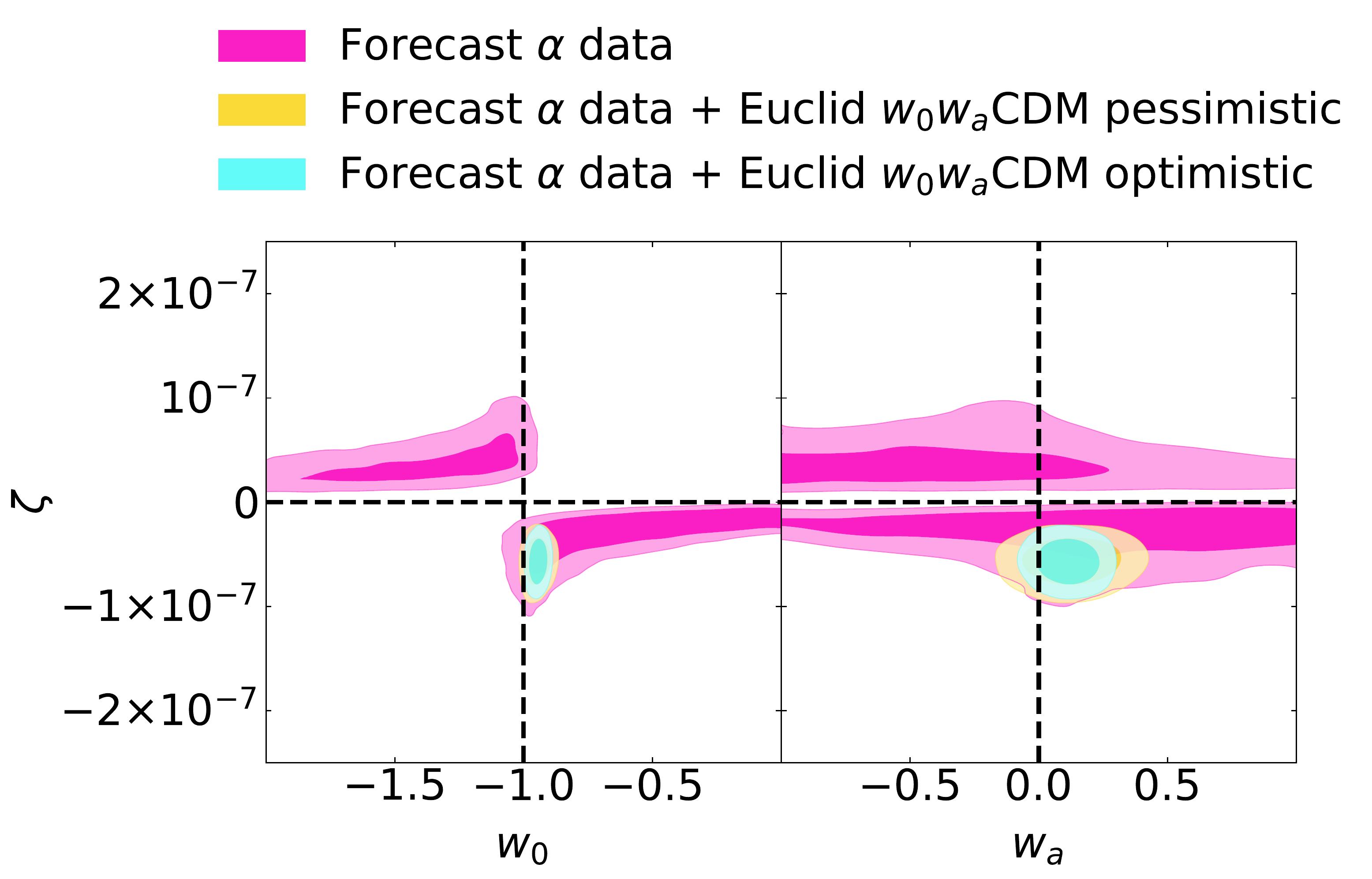}\\
    \includegraphics[width=0.9\columnwidth]{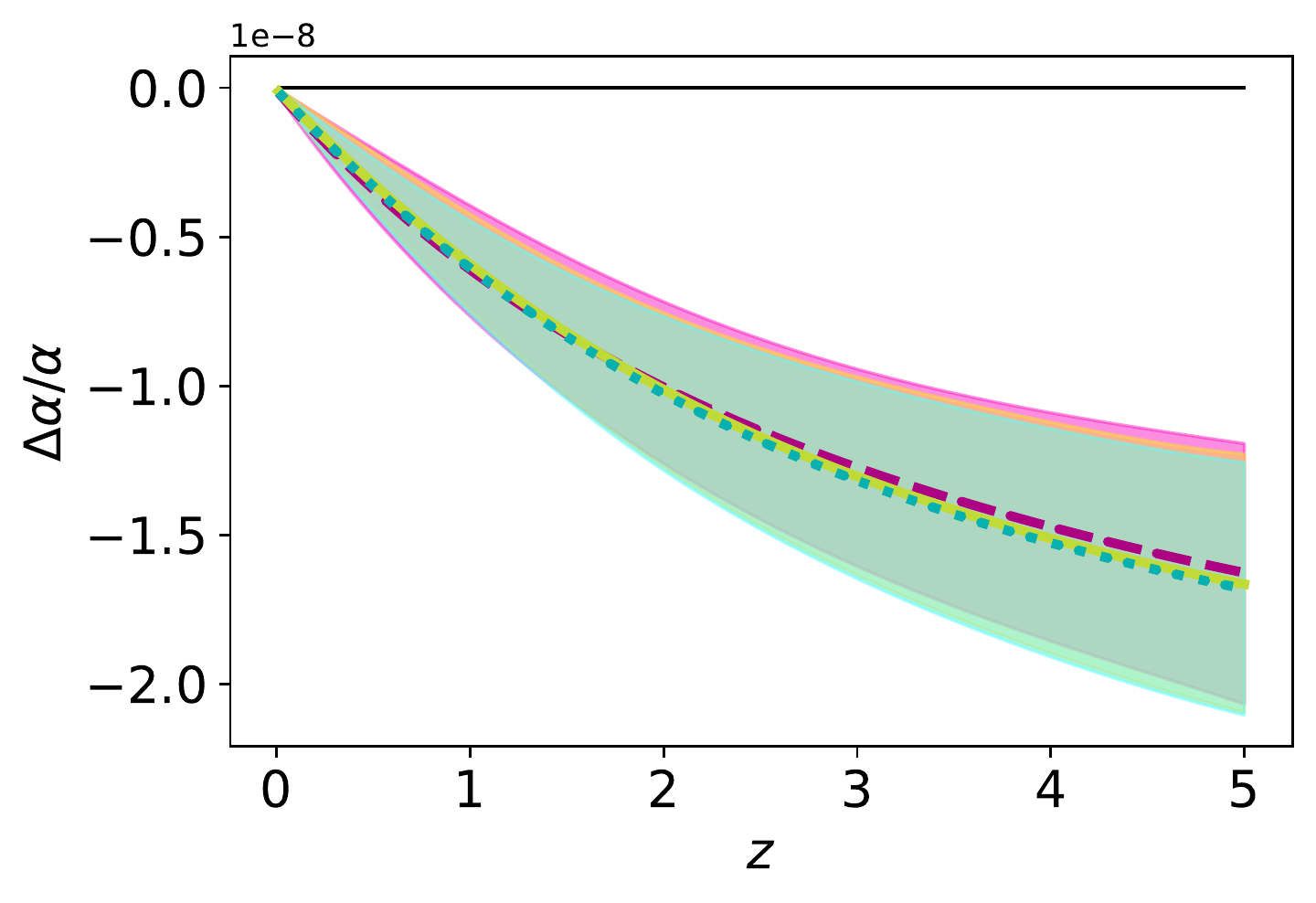}
    \caption{Same as \Cref{fig:ELTLCDM}, but the fiducial cosmology used in this case is the $\zeta w_0w_a$CDM of \Cref{tab:fiducial}, thus a case that deviates from $\Lambda$CDM, with $\zeta=-5\times10^{-8},\ w_0=-0.94,\ w_a=0.1$.}
    \label{fig:ELTnonstd}
\end{figure}

The results shown in \Cref{fig:ELTnonstd} correspond to the $\zeta w_0w_a$CDM fiducial of \Cref{tab:fiducial}. The values chosen for this fiducial make the precise data of ELT incompatible with a vanishing $\Delta\alpha/\alpha$ and this leads to a multimodal posterior distribution for the $\zeta$, $w_0$ and $w_a$ parameters. Looking at \Cref{eq:dalfa,eq:dalfa2}, the symmetry of these peaks with respect to the $(\zeta,w_0,w_a)=(0,-1,0)$ point in the parameter space appears evident, as simultaneously changing the sign of $\zeta$ and $1+w(z)$ leads to the same low-redshift evolution of $\Delta\alpha/\alpha$. Because of this, when \Euclid information is included, we find a breaking of the symmetry between the coupling and CPL parameters, with \Euclid tightening the constraints on $w_0$ and $w_a$ around the fiducial.

Also in this case, as for current $\alpha$ data, the inclusion of the information on $\Delta\alpha/\alpha$ impacts the $(w_0,w_a)$ FoM with respect to what is obtained with \Euclid alone. In the $\Lambda$CDM fiducial, the FoM improves by $26\%$ in both the pessimistic and optimistic cases, while for the $\zeta w_0w_a$CDM fiducial the improvement in the FoM becomes $8\%$.

\begin{table}
\centering
\caption{Mean values and $68\%$ c.l. bounds obtained using forecast $\alpha$ data and their combination with \Euclid forecasts.}
\resizebox{\columnwidth}{!}{
\begin{tabular}{cccc} 
\hline
    \multicolumn{4}{c}{$\Lambda$CDM fiducial} \\
                  & forecast $\alpha$         & +\Euclid pess      &  +\Euclid opt \\
 \hline 
$\Omega_{\rm m}$  & ---                      & $0.3201\pm 0.0037$ & $0.3200\pm 0.0018$\\

$w_0$             & $-1.07\pm 0.48$          & $-0.9998\pm 0.038$ & $-1.000\pm 0.023$\\

$w_a$             & ---                      & $0.00\pm 0.15$     & $0.000\pm 0.085$\\

$H_0$             & ---                      & $67.00\pm 0.33$    & $67.000\pm 0.093$\\

$\zeta\,10^8$  & $0.1\pm 1.9$             & $-0.1\pm 4.8$      & $-0.1\pm 5.9$\\
\hline
     \multicolumn{4}{c}{$\zeta w_0w_a$CDM fiducial} \\
                 & forecast $\alpha$        & +\Euclid pess              &  +\Euclid opt \\
 \hline 
$\Omega_{\rm m}$ & ---                     & $0.3195\pm 0.0034$         & $0.3198\pm 0.0019$\\

$w_0$            & $-0.83^{+0.71}_{-0.30}$ & $-0.946^{+0.029}_{-0.035}$ & $-0.944\pm 0.023$\\

$w_a$            & ---                     & $0.13\pm 0.12$             & $0.113\pm 0.080$\\

$H_0$            & $< 76.1$                & $66.96\pm 0.36$            & $66.993\pm 0.099$\\

$\zeta\,10^8$ & $-0.5^{+4.3}_{-3.1}$    & $-5.7^{+1.6}_{-1.4}$       & $-5.7\pm 1.5$ \\
\hline
\end{tabular} 
}
\label{tab:forres}
\end{table}

\subsection{Bayesian evidence}
The results discussed in this Section make use of Metropolis-Hastings (MH) algorithm to sample the parameter space. This algorithm might fail in reconstructing the posterior shape when this is multimodal. Given the behaviour of some of our posterior distributions we compare the results obtained through MH with those from a nested sampling approach, using the public \texttt{polychord} code \citep{Handley:2015fda,Handley_2015} available in \texttt{Cobaya}, finding compatible results.

Given our use of nested sampling, we obtain as a byproduct of our analysis pipeline an estimate of the Bayesian evidence for each of the cases considered. This allows us to perform a model selection analysis for the different fiducial and experimental settings being considered in this work. We take as reference the $\Lambda$CDM model, thus analyzing all the different data combinations fixing $\zeta=0$, $w_0=-1$, and $w_a=0$. Once the evidence $Z_{\rm ref}$ is computed for this model, we compare it with the evidence $Z$ obtained when these parameters are free to vary. In all the cases considered, we assume the same priors on the cosmological parameters, and therefore their effect on the evidence calculation should cancel out. For the coupling $\zeta$, in the case in which this is free to vary, we always use the MICROSCOPE prior discussed in \Cref{sec:curralpha}.

In \Cref{tab:evidence} we show the difference of the logarithms of the evidence ($K=\log{Z}-\log{Z_{\rm ref}}$); here a positive value indicates a preference for the extended model, while a negative value supports the $\Lambda$CDM model. The results shown highlight how, assuming a $\Lambda$CDM fiducial, the model comparison favours the reference model (negative values) while for $\zeta w_0w_a$CDM the extended model is supported. Thanks to the constraining power of \Euclid, all the cases (pessimistic and optimistic) and for both current and forecast $\alpha$ data provide ``decisive'' evidence for one or the other model according to the Jeffrey's scale \citep[]{Jeffreys:1939xee}. However, it has been noted in \citet[]{Nesseris:2012cq} that the values of the Jeffrey's scale should be interpreted with caution, especially in cases of nested models, as they may lead to biased conclusions.

We note that the extreme values of $K$ shown in \Cref{tab:evidence} are mainly due to the constraining power of \Euclid on $w_0$ and $w_a$. If the reference model is taken to be a $w_0w_a$CDM model, thus with $\zeta=0$, but with $w_0$ and $w_a$ free to vary, the situation changes significantly. In the $\Lambda$CDM fiducial case, the comparison between a model with varying $\alpha$ and the $w_0w_a$CDM model is always inconclusive, as expected since both of them compare similarly with respect to the favoured $\Lambda$CDM model. If we move to the $\zeta w_0w_a$CDM fiducial instead, when using current $\alpha$ data, the comparison of the two models is still inconclusive, thus highlighting how the decisive preference with the previous reference was totally driven by \Euclid constraints on CPL parameters, and that the sensitivity of current $\alpha$ data do not allow to distinguish the model under examination here from a simple $w_0w_a$CDM. The case of forecast data combined with \Euclid instead still shows a decisive evidence in favour of the varying $\alpha$ model, also with respect to the $w_0w_a$CDM reference. This clearly shows how the improvement brought by the combination of ELT and \Euclid is able to distinguish a varying $\alpha$ model not only from the standard $\Lambda$CDM case, but also from a cosmology where the dark energy scalar field is not coupled to the electromagnetic sector.

\begin{table}
\centering
\caption{Differences ($K$) in the logarithm of the Bayesian evidence, between the extended model allowing for $\alpha$ variation and the reference $\Lambda$CDM model. The cases considered here only show the combination of \Euclid and $\alpha$ data}
\resizebox{\columnwidth}{!}{
\begin{tabular}{c|cc|cc} 
\hline
                  &  \multicolumn{4}{c}{$\Lambda$CDM fiducial}  \\
                  &  \multicolumn{2}{c|}{$\Lambda$CDM reference} & \multicolumn{2}{c}{$w_0w_a$CDM reference}\\
                  & pessimistic    & optimistic & pessimistic    & optimistic\\
 \hline 
current $\alpha$  & $-5.8$         & $-7.0$     & $0.01$         & $-0.27$\\

forecast $\alpha$ & $-6.4$         & $-7.4$     & $-0.65$        & $-0.92$ \\
\hline
                  &  \multicolumn{4}{c}{$\zeta w_0w_a$CDM fiducial}  \\
                  &  \multicolumn{2}{c|}{$\Lambda$CDM reference} & \multicolumn{2}{c}{$w_0w_a$CDM reference}\\
                  & pessimistic    & optimistic & pessimistic    & optimistic\\
 \hline 
current $\alpha$  & $9.5$          & $48.8$     & $-0.32$        & $-0.50$ \\

forecast $\alpha$ & $15.6$         & $55.1$     & $5.96$         & $5.71$ \\
\hline
\end{tabular} 
}
\label{tab:evidence}
\end{table}

\section{Coupling null test results \label{sec:GA}}

In this Section we present the GA reconstruction of the coupling $\zeta$ as a null test of whether any possible redshift dependence of $\alpha$ could be detected through the combination of \Euclid and current astrophysical data or  future ELT data. We use a machine learning approach based on the GA to reconstruct $\zeta$, as this effectively provides a null test for the constancy of the gauge kinetic term and its linear expansion given by \Cref{eq:linearexp}. 

\subsection{Euclid and current $\alpha$ data results}

\begin{figure}
\centering
\includegraphics[width=0.9\columnwidth]{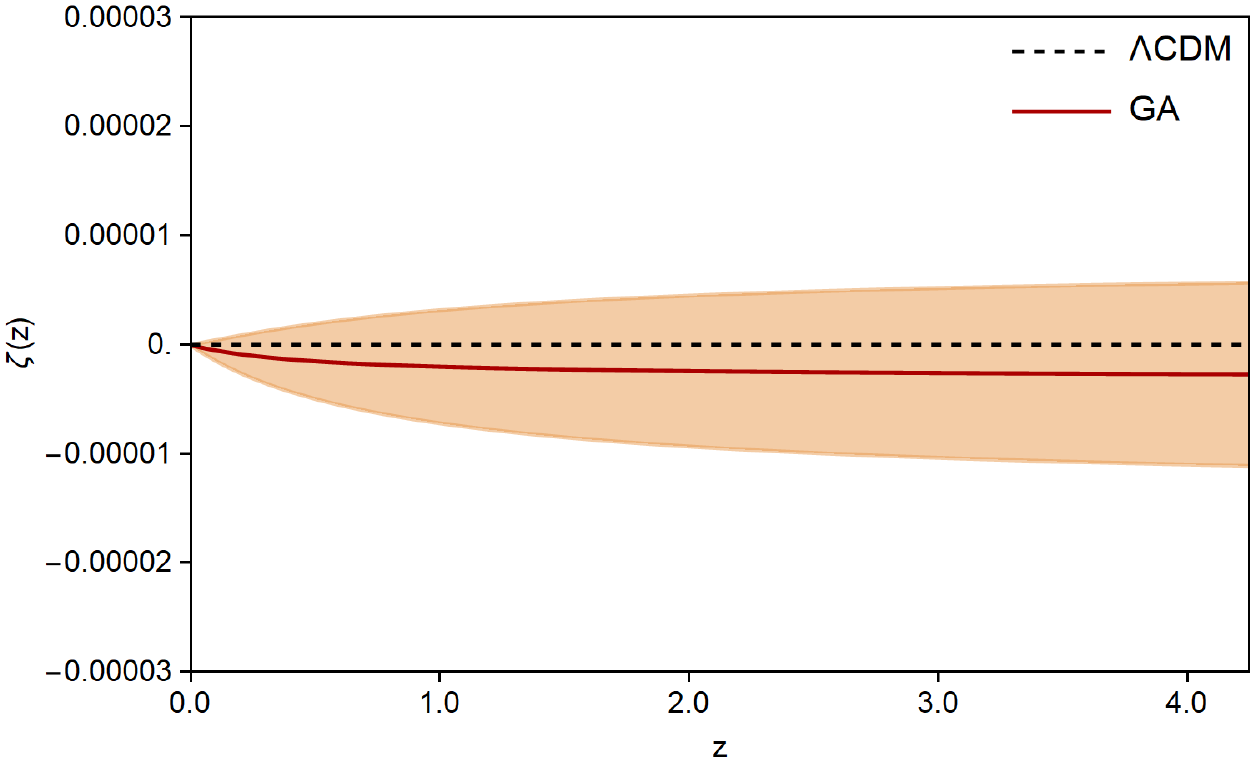}
\includegraphics[width=0.9\columnwidth]{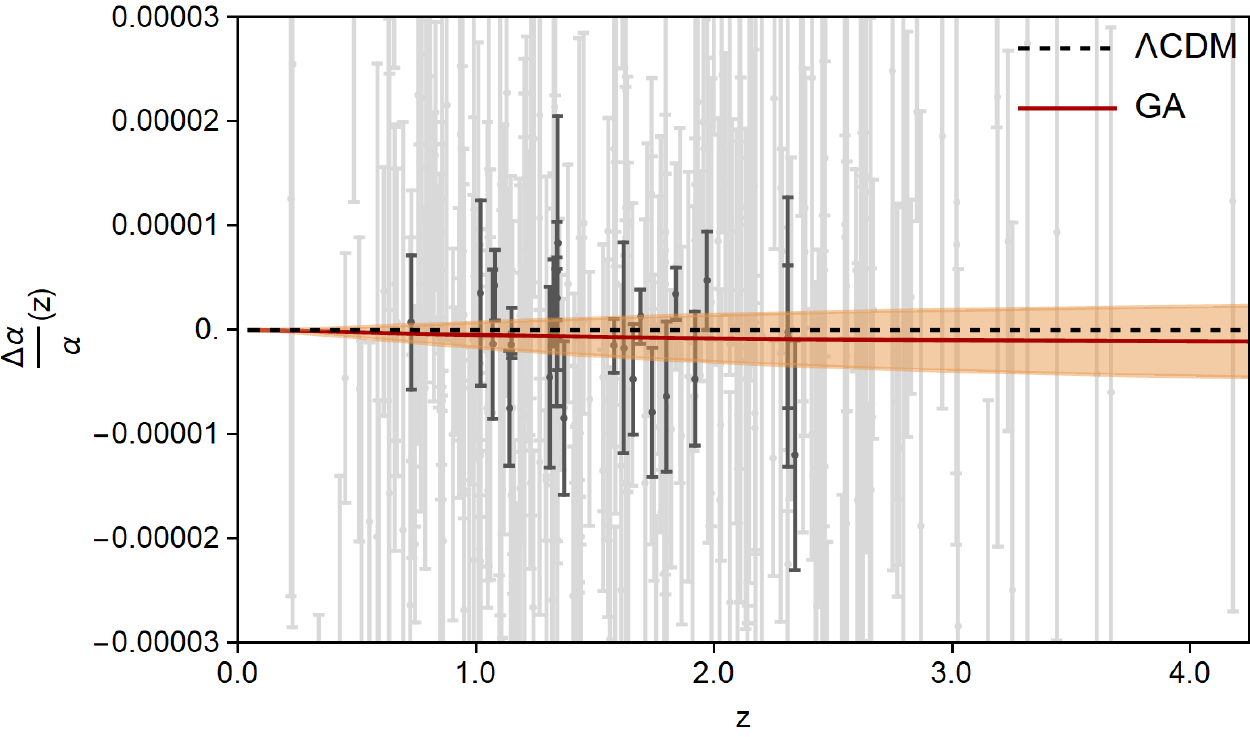}
\caption{{\it Top panel}: GA Reconstruction of the coupling $\zeta$ as a function of redshift using the currently available $\alpha$ measurements and \Euclid forecast constraints with a $\zeta w_0w_a$CDM fiducial. {\it Bottom panel}: Reconstruction of the relative variation of the fine structure constant $\Delta\alpha/\alpha$. The Webb data is shown in grey background points and the dedicated $\alpha$ measurements are shown in black background points. In both panels the red line corresponds to the GA reconstruction, while the shaded region is the $1\sigma$ error.}
    \label{fig:GA1}
\end{figure}

First, we directly reconstruct the coupling $\zeta$ using the currently available $\alpha$ measurements, comprising the combination of Webb archival data and the dedicated $\alpha$ measurements, atomic clocks constraints and the MICROSCOPE bound, together with the \Euclid forecast constraints with a $\zeta w_0w_a$CDM fiducial. In the top panel of \Cref{fig:GA1} we show the GA reconstruction of the coupling $\zeta$ as a function of redshift (red line) together with the $1\,\sigma$ error shaded region.

The nominal GA reconstruction of $\zeta(z)$ is found to be 
\be
\zeta_\textrm{GA}(z)=-\frac{0.011+z\;(1.558+3.041 z)}{0.256+z (1.011+z)} 10^{-6},
\ee 
but as can be seen in the Figure, $\zeta$ is fully consistent with zero within the errors. At high redshifts the GA leads to a value for the coupling  of $\zeta_\textrm{GA}(z\sim 4)\simeq(-2.704\pm 8.293) \times 10^{-6}$. On the other hand, at $z=0$ the GA gives the value $\zeta_\textrm{GA}(z=0)=(-4.285\pm 1.510) \times 10^{-8}$, which is in very good agreement with the parametric case when the coupling is assumed to be constant. 

For completeness, in the bottom panel of \Cref{fig:GA1} we show a  reconstruction of the relative variation of the fine structure constant $\Delta\alpha/\alpha$  (red line), under the previously mentioned assumptions, in order to assess its redshift trend. Here the Webb archival data is shown in grey background points, while the dedicated $\alpha$ measurements are shown  in black background points. We confirm that the allowed  $\Delta\alpha/\alpha$ is tightly constrained around zero, even at high redshifts.

\subsection{Synergy between \Euclid and ELT}

\begin{figure}
\centering
\includegraphics[width=0.9\columnwidth]{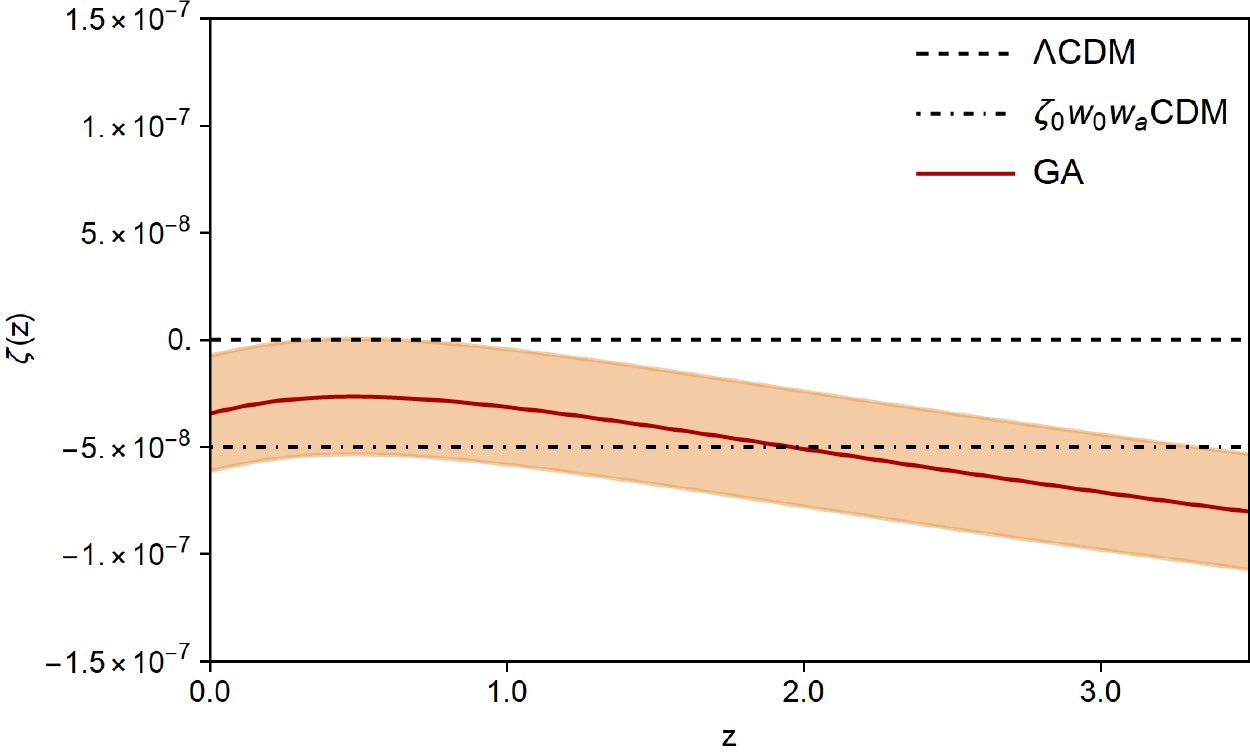}
\includegraphics[width=0.9\columnwidth]{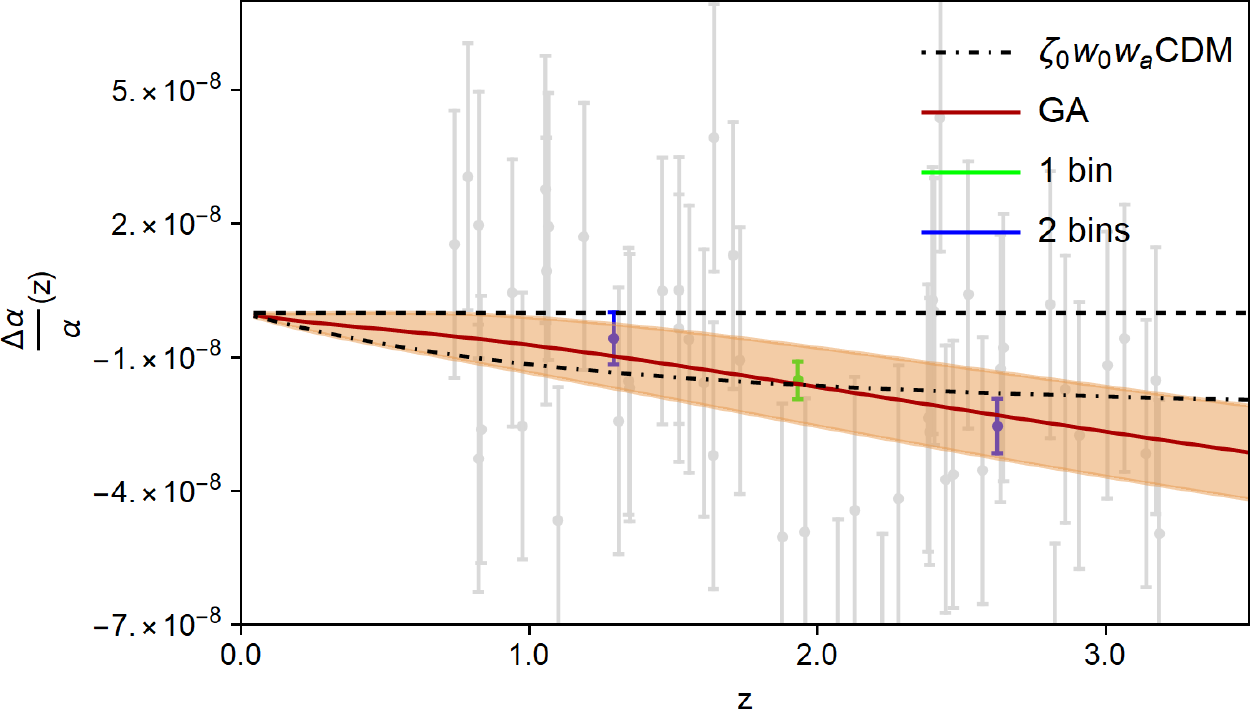}
\caption{{\it Top panel}: GA reconstruction of the coupling $\zeta$ as a function of redshift (red line) using the forecast ELT $\alpha$ data (grey background points), the atomic clocks constraints and the MICROSCOPE bound (both at $z=0$) and \Euclid forecast constraints with a $\zeta w_0w_a$CDM fiducial. {\it Bottom panel}: Reconstruction of the relative variation of the fine structure constant $\Delta\alpha/\alpha$ (red line) along with binned values of the data in one (green point) and two bins (blue points). In both cases the red line corresponds to the GA reconstruction, while the shaded region is the $1\sigma$ error.\label{fig:GA3}} 
\end{figure}

Next, we consider the scenario of combining \Euclid data with the forecast ELT data, in the case where the fiducial model has a non-zero coupling. In the top panel of \Cref{fig:GA3} we show the GA reconstruction of the coupling $\zeta$ as a function of redshift (red line) using the forecast ELT $\alpha$ measurements, together with the \Euclid forecast constraints with a $\zeta w_0w_a$CDM fiducial. We use the same atomic clocks constraints and the MICROSCOPE as in the previous sub-section. We find that the GA reconstruction is consistent with the fiducial value (shown with the dot-dashed line) within the $1\sigma$ bound. In particular, at high redshifts $(z>1)$ a deviation from zero is detected, in agreement with the fiducial value used in our mocks, highlighting the importance of the extension of the redshift lever arm provided by the ELT measurements of $\alpha$.

In the bottom panel of \Cref{fig:GA3} we also show a GA reconstruction of the relative variation of the fine structure constant $\Delta\alpha/\alpha$ (red line). As a test of our approach and given the much higher sensitivity of the ELT data, we also bin the data either in one bin (green point) or in two bins (blue points) splitting the ELT data at $z=2$. We find that the GA reconstruction of the relative variation of the fine structure constant is in excellent agreement with both the fiducial model, given by the dot-dashed line, and with the binning of the data.

\section{Discussion and outlook}\label{sec:summary}

When studying theoretical models beyond the concordance $\Lambda$CDM framework, many realistic extensions introduce couplings between the different degrees of freedom. In particular, the coupling of physically realistic dynamical dark energy scalar fields to the electromagnetic sector will lead to a time dependence of the fine-structure constant \citep{Carroll,Dvali,Chiba}, which could be detected and interpreted as a smoking gun for the existence of extra scalar fields. In this work we have studied the role of the forthcoming \Euclid mission in constraining such theoretical models. \Euclid will provide us with very precise cosmological information using the clustering of galaxies and the weak lensing measurements from the large-scale structure of the Universe. However, these probes are not enough to constrain the full parameter space of these models. We therefore need to add astrophysical (and local) data, specifically to constrain the coupling between the dark energy scalar field and the electromagnetic sector.

In this work we have considered current astrophysical tests of the stability of the fine-structure constant from quasi-stellar object spectral lines (both from archival data and from dedicated measurements), as well as current laboratory constraints on the present-day drift rate of $\alpha$ from atomic clock experiments and constraints on the E\"otv\"os parameter from the MICROSCOPE satellite. However, at the time \Euclid data will be available we expect to have even more precise astrophysical measurements of $\alpha$, so we have also forecast the precision of the high-resolution ultra-stable spectrograph HIRES at the Extremely Large Telescope. We have used both a parametric approach, under a standard likelihood analysis, and a non-parametric machine learning class of stochastic optimization methods to forecast the constraints from the joint analysis of \Euclid and astrophysical and local data.

Starting with the parametric approach, we have first considered the synergy between \Euclid and current measurements of the relative variation of the fine-structure constant. Our baseline scenario has a $\Lambda$CDM fiducial, meaning that the fiducial corresponds to a vanishing coupling constant and a cosmological constant as dark energy. In \Cref{sec:likecurr} we have seen that \Euclid very significantly restricts the allowed values of the fine-structure constant as a function of redshift compared to the constraints with astrophysical and laboratory data alone. This is due to \Euclid's ability to constrain the dark energy equation of state parameters. However, for this same reason, and because of the degeneracy between the coupling constant and $w_0$ and $w_a$ when they are close to the cosmological constant corresponding values, the addition of \Euclid data loosens the constraints on the coupling $\zeta$ compared to the constraints from fine-structure constant data alone.

We have then performed the same analysis using the $\zeta w_0 w_a$CDM fiducial for the \Euclid results, where the coupling fiducial is  non-null and the fiducial values of $w_0$ and $w_a$ no longer correspond to a cosmological constant. In this case the increase of constraining power on the evolution of $\alpha$ when adding \Euclid data is still present. Concerning the bounds on the coupling constant, the addition of \Euclid data only marginally loosens the constraints, since the fiducial is not exactly on $\Lambda$CDM and the degeneracy between the coupling constant and the dark energy parameters is partially broken. It is also worth mentioning that adding these astrophysical and local tests of the variation of the fine-structure constant improves the constraints on the dark energy equation of state parameters from \Euclid data alone, with the FoM for dark energy parameters improving between 3\% and 18\%, as long as the model considered connects the variation of $\alpha$ to the dark energy parameters, thus making the $\alpha$ data sensitive to them.

Still with the parametric approach, we have considered the synergy between \Euclid and the ELT. The results obtained have been qualitatively the same, with \Euclid data helping to constrain the evolution of the fine-structure constant as a function of redshift while the astrophysical and local data help constraining the dark energy equation of state. However, given the fact that the ELT data is much more precise than current measurements, the contribution of \Euclid on $\Delta \alpha/\alpha$ is proportionally somewhat smaller, while the contribution of ELT on $w_0$ and $w_a$ is slightly larger, with the FoM now improving between $8\%$ and $26\%$. Furthermore, we also performed a model comparison analysis, which highlighted how both current and forecast $\alpha$ data, in combination with \Euclid, can significantly distinguish between $\Lambda$CDM and a model with a coupled scalar field, while only when considering ELT forecast can the latter be distinguished from a $w_0w_a$CDM cosmology with a vanishing $\zeta$ coupling.

We have also used a model-independent approach to reconstruct the coupling between the dark energy scalar field and the electromagnetic sector. This is effectively a null test reconstruction of the behaviour of the coupling $\zeta$. Specifically, any deviation from a constant coupling would either suggest unidentified systematics in the astrophysical data or indicate that the assumptions made in \Cref{sec:theory} break down and our modelling is not accurate enough to explain the observations; in the latter case this would imply that the putative dynamical dark energy and varying $\alpha$ would not be due to the same underlying physical mechanism -- which would in itself be a significant result. Our analysis shows how the GA are able to reconstruct the coupling function in agreement with the fiducial values assumed, and are compatible with a constant coupling.

Overall, we have found that the synergies between the main probes of \Euclid and astrophysical measurements of $\alpha$ can tightly constrain models where the dark energy scalar field is coupled to the electromagnetic sector. In addition to this one must notice that other \Euclid probes (such as a possible SNIa survey) can also be directly sensitive to a varying fine structure constant, and this would further improve the contribution of \Euclid on tests of such coupled models.

\appendix

\section{Current data compatibility}\label{sec:app_current}

As pointed out in \citet{ROPP,Meritxell}, the currently available astrophysical tests of the stability of the fine structure constant are in slight tension with each other; the weighted mean of $\Delta\alpha/\alpha$ obtained through the Webb archival data is in fact in tension of $\approx2\sigma$ with the one obtained through the recent dedicated measurements. In order to be able to combine these datasets, as we did throughout our paper, we must assess the significance of such tension.

In the parametric approach, we can estimate the concordance of the datasets by computing the Bayesian ratio
\begin{equation}
    K = \frac{Z({\rm Webb+recent})}{Z({\rm Webb})Z({\rm recent})}\, ,
\end{equation}
where $Z({\rm Webb})$ is the evidence when using Webb data alone, $Z({\rm recent})$ when only recent data are considered and $Z({\rm Webb+recent})$ the case when the two are considered in combination. Such a Bayesian ratio can be used for model comparison between the case in which the two datasets are used to fit the same set of cosmological parameter, $Z({\rm Webb+recent})$, and the case in which these might differ for the two datasets, $Z({\rm Webb})Z({\rm recent})$.

We obtain these values using \texttt{polychord} to sample our free parameters and, when considering only astrophysical $\alpha$ measurements, we find that $\ln{K}\approx0.2$, a value that is not able to provide any conclusive evidence for either the concordance or discordance of the data. When instead we include the information brought by local measurements, we obtain $\ln{K}\approx1.6$, thus an improved significance toward the concordance of the data. Such an increase in the value of $\ln{K}$ comes from the fact that the local measurements dominate the constraints obtained through the parametric approach, thus hiding any possible discordance of the two sets of astrophysical data.

This does not apply in our null test reconstruction of $\zeta(z)$; here, the local measurements only contribute to the very low redshift reconstruction, while astrophysical data dominate at higher redshift. Because of this tension between the Webb archival data and the recent dedicated measurement, the reconstruction performed with the GA attempts to be in agreement with both datasets, and this can potentially increase the error on the reconstructed function. One thus expects that removing one of the datasets, the error on the reconstruction will decrease.

\begin{figure}
\centering
\includegraphics[width=\columnwidth]{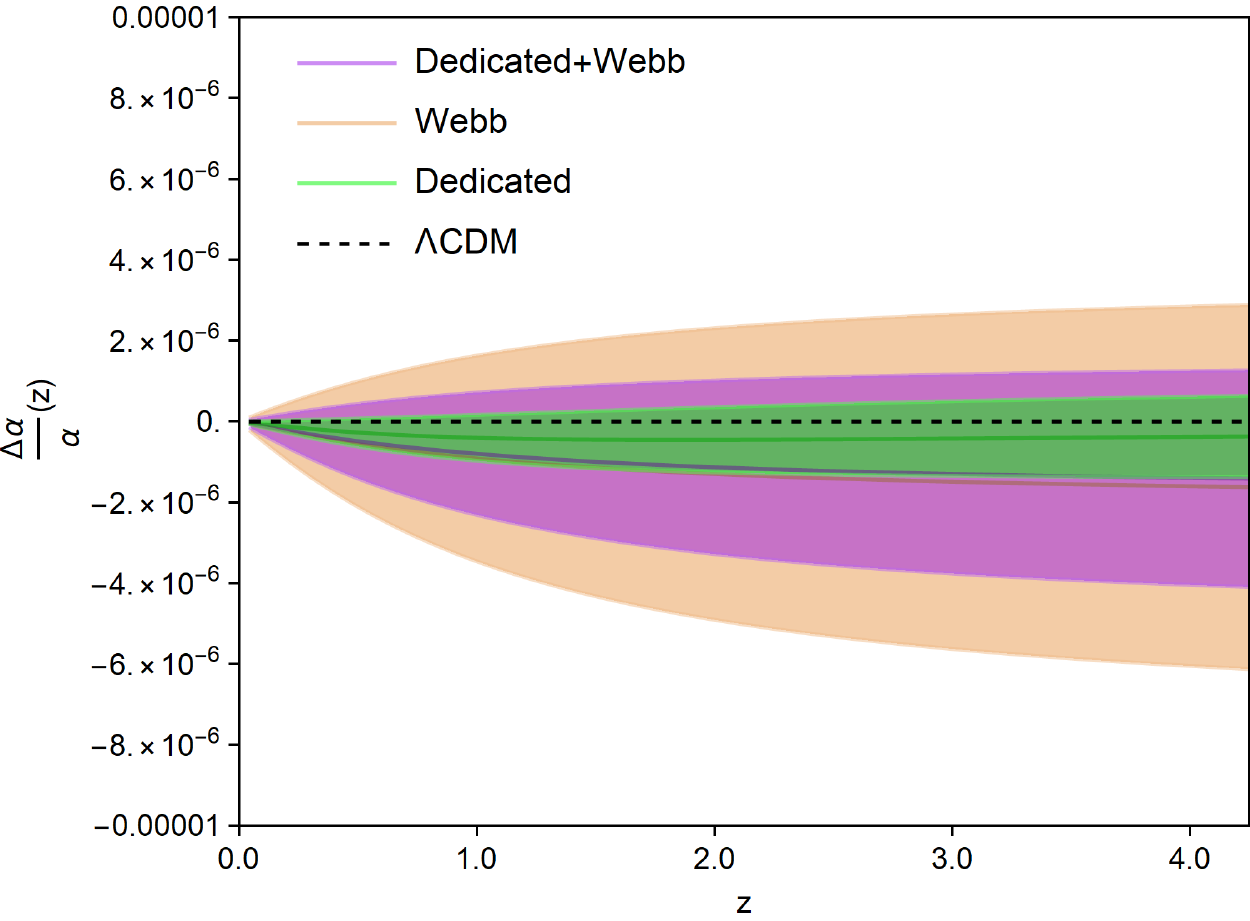}
\caption{GA reconstruction of the relative variation of the fine structure constant $\Delta\alpha/\alpha$ as a function of redshift. The green line corresponds to the dedicated measurements, the orange line to the Webb archival data and the magenta line to the combination of the two, while in all cases the shaded region is the $1\sigma$ error.\label{fig:GA2}}
\end{figure}

In particular, in \Cref{fig:GA2} we show the GA reconstruction of the relative variation of the fine structure constant $\Delta\alpha/\alpha$ as a function of redshift. The green line corresponds to the dedicated measurements, the orange line to the Webb archival data and the magenta line to the combination of the two, while in all cases the shaded region is the $1\sigma$ error. As expected, the reconstruction of the dedicated measurements has a smaller error (green shaded region), but when we combine them with the Webb archival data then the combined error region not only does not decrease, but instead increases due to the tension, as observed in the magenta shaded region. Equivalently, we see that when we remove the Webb data from the combination of the two, counter-intuitively the error of the reconstruction decreases. However, overall the reconstructions are still compatible with each other and with zero. 

\begin{acknowledgements}
MM has received the support of a fellowship from ``la Caixa'' Foundation (ID 100010434), with fellowship code LCF/BQ/PI19/11690015, and the support of the Centro de Excelencia Severo Ochoa Program SEV-2016-059. The work of CJM was financed by FEDER -- Fundo Europeu de Desenvolvimento Regional funds through the COMPETE 2020 -- Operational Programme for Competitiveness and Internationalisation (POCI), and by Portuguese funds through FCT - Funda\c c\~ao para a Ci\^encia e a Tecnologia in the framework of the project POCI-01-0145-FEDER-028987. IT acknowledges support from the Spanish Ministry of Science, Innovation and Universities through grant ESP2017-89838, and the H2020 programme of the European Commission through grant 776247. SN acknowledges support from the research project  PGC2018-094773-B-C32, the Centro de Excelencia Severo Ochoa Program SEV-2016-059 and the Ram\'{o}n y Cajal program through Grant No. RYC-2014-15843.

\AckEC
\end{acknowledgements}

\bibliographystyle{aa}
\bibliography{references}

\end{document}